\documentclass{mn2e} 
\usepackage{amssymb}
\input epsf.sty 
\newif\ifAMStwofonts 
\AMStwofontstrue 
 
\begin{document} 
 
\title{Variability of accretion flow in the core of the Seyfert galaxy NGC 4151} 
\author[Czerny et al.] 
{B. Czerny$^1$,  V.T. Doroshenko$^2$, M. Niko\l ajuk$^1$, A. Schwarzenberg-Czerny$^{1,3}$,  
\newauthor
Z. Loska$^1$ and G. Madejski$^4$\\ 
 $^1$N. Copernicus Astronomical Center, Bartycka 18, 00-716 Warsaw, Poland\\ 
 $^2$Crimean Laboratory of Sternberg Astronomical Institute, Nauchny, 
    Crimea, 98409 Ukraine \\
  $^3$Astronomical Observatory of Adam Mickiewicz University, 
    ul. S\l oneczna 36, 60-286 Pozna\' n, Poland \\ 
  $^4$Stanford Linear Accelerator Center, 2575 Sand Hill Road, Menlo Park, 
    CA 94025, USA} 
%\date{Accepted 1997 . 
%      Received 1997 ; 
%      in original form 1997 } 
%\pubyear{1997} 
\maketitle 
\begin{abstract} 
We analyze observations of the Seyfert galaxy NGC~4151 covering
90 years in the optical band and 27 years in the 2 - 10 keV X-ray 
band. We compute the Normalized Power Spectrum Density (NPSD), the 
Structure Function (SF) and the Autocorrelation Function (ACF) for 
these data.  The results show that the optical and X-ray variability 
properties are significantly different. X-ray variations are 
predominantly in the timescale range of 5 - 1000 days. The optical  
variations have also a short timescale component which may be related 
to X-ray variability but the dominant effect is the long timescale 
variability, with timescales longer than $\sim 10$ years. We compare 
our results with observations of NGC~5548 and Cyg X-1. We conclude 
that the long timescale variability may be caused by radiation 
pressure instability in the accretion disk, although the observed 
timescale in NGC~4151 is by a factor of few longer than expected.  
X-ray variability of this source is very similar to what is observed 
in Cyg X-1 but scaled with the mass of the black hole, which suggests 
that the radiation pressure instability does not affect considerably 
the X-ray production. 
 
\end{abstract} 
 
 
\section{Introduction}\label{cos}
 
NGC~4151 ($z=0.00332$; $D=13.2$ Mpc assuming Hubble constant 
75 km s$^{-1}$ Mpc$^{-1}$) is one of the classical Seyfert 1 galaxies. 
It is among the most frequently observed AGN and therefore it is an 
excellent object for variability studies, although it is both typical and 
atypical representative of its class, as nicely and extensively summarized 
by Ulrich (2000). We therefore chose it as a second source (after 
NGC~5548; Czerny, Schwarzenberg-Czerny, \& Loska 1999) for which the 
nature of the variability can be studied by computing the normalized power 
spectrum density (hereafter NPSD) in both {\it the optical and the 
X-ray bands}.

The host galaxy is a typical spiral galaxy SAB(rs)ab of 11.3 mag
brightness in B band (Perez et al. 1998). The extinction due to our 
Galaxy is low (A$_V<0.01$ in B band) because NGC~4151 is located
near the North Galactic Pole ($b_{II}=75.1^{\circ}$).  However, the 
intrinsic extinction in NGC~4151 is considerable:  in the optical region, 
A$_V=1^m.0$ in BLR and A$_V=0^m.7$ in NLR (Ward et al. 1987). Direct 
estimates in the UV bands give much less extinction (E$_{B-V}=0.06$) 
for wavelength $\lambda 2200$ \AA\, corresponding to A$_V=0^m.18$ 
(Wu \& Weedman 1978). The hydrogen column density inferred from the 
equivalent widths of ultraviolet absorption CIII lines is about 
$1.7 \times 10^{19}$ cm$^{-2}$ (Kriss et al. 1992), smaller than the 
inferred X-ray column density.  The X-ray observations clearly indicate 
that the nuclear X-ray continuum is significantly obscured, by an 
equivalent hydrogen column $N_H = (1-10) \times 10^{22}$ cm$^{-2}$) 
(Yaqoob et al. 1993), which would correspond to A$_V\approx5^m - 50^m$ 
according to the standard relation between A$_V$ and N$_H$ from 
Reina \& Tarenghi (1973).

HST observations suggested that the line of sight towards the nucleus
(at least in the epoch when the observation was performed, on 1991
June 18/19) lies outside the ionization cone (Evans et al. 1993). 
However, broad emission lines - which can vary in strength - 
are usually easily detected, although with superimposed absorption 
features (e.g.  Brandt et al. 2001); only during one observation, 
in 1984, the broad component of H$_{\beta}$ was absent (Lyutyj, 
Oknyanskij \& Chuvaev 1984; Penston \& P\'erez 1984; see also Pronik, 
Sergeev, \& Sergeeva 2001) and the spectrum more closely resembled 
that of a narrow emission line or a Seyfert 2 type galaxy.  Therefore 
it seems that most of time the nucleus is viewed directly and the scattered
component characteristic for Seyfert 2 galaxies does not dominate the
spectrum, apart from the soft X-ray band (see Ogle et al. 2000). 

Regardless of the fact that X-ray extinction is high, the intrinsic
X-ray spectrum seems to be fairly typical for a Seyfert galaxy
(Zdziarski, Johnson, \& Magdziarz 1996).  The brightness of the 
nucleus varies dramatically on all timescales in all energy bands. 
Spectrum and variability in the 50 - 150 keV band was studied by 
Johnson et al. (1997);  the source shows a factor of 2 variability 
in days and years. The X ray (2 - 10 keV) variability of this source 
was studied 
in more detail using Fourier analysis, starting with the paper by Fiore 
et al. (1989).  It was found (Papadakis 
\& McHardy 1995) that X-ray variability power spectrum of NGC~4151 
covering the range $10^{-6} - 10^{-8}$ Hz can be described by a power law, 
which is significantly flatter than that measured on short time 
scales ($10^{-2} - 10^{-5}$ Hz) derived from EXOSAT observations.  
The power law index measured by Hayashida et al. (1998) on the basis 
of the Ginga data in the frequency region of $10^{-4} - 10^{-6}$ Hz has 
an index $\alpha = 2.1$.

The long-term optical variability of the continuum in the UBV band 
was studied by Lyuty \& Doroshenko (1999). They showed that a long photometric 
minimum in years 1984 - 1989 separates two active phases: 1968 - 1988 
(Cycle $\cal A$) and 1989 - 2000 (Cycle $\cal B$). The decrease of 
brightness continued in 2000 and at the end of 2000, the brightness of the 
variable component reached almost the same level as in 1987 - 1988
(Doroshenko et al. 2001). These phases are also seen in the JHKL-bands
(Lyutyi, Taranova, \& Shenavrin 1998).  

The power spectrum density (PSD) of the NGC~4151 optical flux variations 
in 1968 - 1987 was studied by Terebizh, Terebizh, \& Biryukov (1989). 
It was found that the PSD in the frequency range 
$10^{-4} - 2 \times 10^{-2}$ day$^{-1}$ (time interval from 50 days 
to 30 years) corresponds to flicker noise with slope $\approx$ 1 in 
logarithmic scales and the light curve may be interpreted as a result 
of superposition of flares randomly distributed in time.  The behavior 
of the structure function (hereafter SF), focusing on the intra-night 
variability and conducted between 1989 - 1996 was studied by Merkulova, 
Metik \& Pronik (2001).  Collier \& Peterson (2001), using the
same technique, determined the characteristic timescale to be 
$13^{+11}_{-5}$ days. 

Long timescale trends in optical variability of NGC~4151 were studied
by Lyutyj \& Oknyanskij (1987) and  Fan \& Su (1999), on the basis of data
covering the period from 1910.  Lyutyj \& 
Oknyanskij (1987) noted the quasi-periodic changes with typical time 
of 4 and 14 years. Also Fan \& Su (1999) claimed the presence of the 
periodicity, with the period 14.0 $\pm$ 0.8 yr, during which both 
the brightness and the spectral type changed.  It is not surprising 
that both groups found the similar quasi-periodical processes in 
optical light curve because the main data set used by both teams was 
the same. Some earlier papers also claimed the presence of the 
periodic variability albeit with other values of the period (e.g. 
Pacholczyk (1971) found a period of 5.1 years), while other papers 
argued against any strict periodicity (e.g. Pacholczyk et al. 1983;  
Terebizh et al. 1989).   

The short term multiwavelength variability was best studied by AGN 
Watch team and was summarized by Edelson et al. (1996). They concluded 
that significant and correlated variability was observed in the X-ray, 
UV and optical bands, with phase differences consistent with zero 
lag and normalized variability amplitude decreasing with 
increasing wavelength.  They calculated the power spectrum in the UV 
on timescales of $\approx$ 0.2 - 5 days and showed that the power 
spectrum is falling down rapidly at short timescales and the bulk 
of the variability power is on timescales of days or longer. 
 
In the present paper we reanalyze the optical and X-ray data 
available in the literature which cover respectively 90 and 25 years.  
Such a broad dynamical range allows us to draw conclusions about 
the character of the accretion flow and the possible nature of the 
instabilities responsible for the observed variability. 

\section{Observational data}\label{s2}

\subsection{Optical band}\label{s21}

The observational data in optical region used in the present paper 
come from different sources. 
Long data sets were collected by the team at the Crimean Station of
the Sternberg Astronomical Institute (1968 - 2000), the Special
Astrophysical Observatory in the Caucasus (1997 - 2000), and the
Maidanak Observatory of the Ulugbek Astronomical Institute in
Uzbekistan (1990 - 2000).  Below, we will refer to these data as 
the Crimean data. The data collected in the three bands (U, B, 
and V) cover the period from 1968 till 2000.  The data points are not
distributed uniformly since the source was at some seasons unobservable 
at the sites involved.  Nonetheless, the sets consist of about 1150
points in each color, thus providing an exceptionally long and well
sampled light curve for any AGN. The light curve and the discussion of
the variability amplitude and color changes were presented by Lyuty \&
Doroshenko (1999) and Doroshenko et al. (2001). Additionally, shorter
data set (about 300 points during 1989 - 1999) in the infrared band R is
also available (Doroshenko et al. 2001).

Shorter, but more densely spaced data set was taken from the AGN Watch
team (Kaspi et al. 1996). Those data cover only 98 days but they add 62 
points to the long set. Continuum flux in the spectral band 4560 - 4640
\AA\ was fitted to the flux in the B-band of Crimean data set using
regression relation based on 17 common dates of observations.

The majority of the oldest data points was derived by Pacholczyk et al. 
(1983) from the old Harvard and Stewart Observatories 
photographic plates. These data 
cover the period from 1910 until 1968. Besides those observations we 
used data from Fitch, Pacholczyk, \& Weymann (1967), Cannon, Penston,
\& Brett (1971), Penston et al. (1971), and Oknyanskij (1978; 1983).  
Comparison of the photographic magnitudes between various data sets 
and with Crimean photoelectric data in B-band for overlapping dates of 
observations allowed us to reduce the photographic data into photoelectric 
B-bands magnitudes. 

As a result, in the B-band we obtained a quite long data set covering years 
1910 - 2000. We must note that the similar work of gathering all
photographic observations was made by Lyuty and Oknyanskij (1987), but
in comparison with this work we added more photoelectric observations made 
after 1984. A subsequent compilation of the 1910 - 1991 data was prepared by
Longo et al. (1996) but they used even older data from heterogeneous 
sources and interpolated the data in order to obtain equally spaced
data points which lead to large errors so finally they had to ignore
the first 30 years of observations. 

\begin{figure} 
\epsfxsize = 90 mm 
%\epsfbox{fig1_tot.eps} 
%\epsfbox{LC-B.PS} 
%\epsfbox{Lc-b.ps}
%\epsfbox{from_valya/f1-lcb.ps} 
\epsfbox{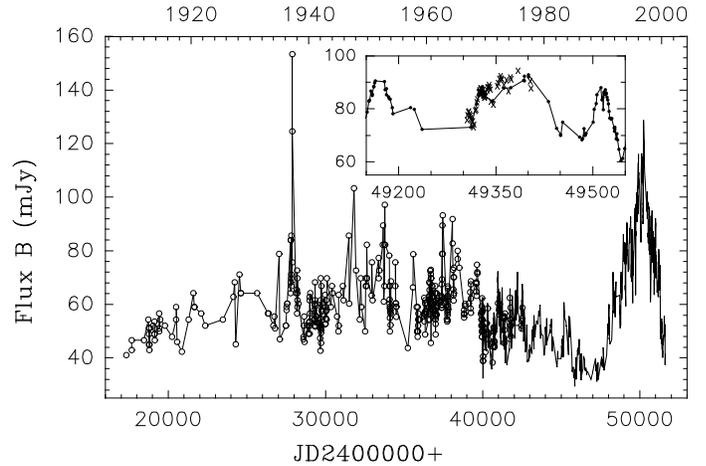}
\caption{The light curve of NGC~4151 in the B band 
from the historical plate data (open circles connected with a 
continuous line) and Crimean 
observations (solid line). Inserted expanded fragment shows the AGN Watch data
(crosses) together with Crimean data (continuous line). The errors are on 
the order of 0.2 mag ($\sim 20$ \%) for the historical plate data and 
$\sim 1.5$ \% for the Crimean observations. 
\label{fig:Bcurve}}
\end{figure}

The full combined light curve of the NGC~4151 nucleus in the B band is 
shown in Fig.~{\ref{fig:Bcurve}}.  Open circles mark the historical 
photographic plate data while the Crimean data are shown with the 
continuous line. A small expanded fraction of the light curve 
from JD2449150 till JD2449550 is inserted in that Figure.  
It includes the three month period when the dense AGN Watch data were 
collected. Crimean data are shown with continuous line and
the AGN Watch points are marked with crosses.  The figure demonstrates 
that the variability in 1910 - 1978 based on photographic data is 
similar in its character to the variability in 1968 - 1989 covered by 
photoelectric observations. Also the AGN Watch data agrees well with 
the corresponding fragment of the Crimean observations:  
the AGN Watch data points are located on the quasi-linear 
part of the light curve showing local rise in brightness. Therefore, we
consider the old data as generally reliable.

The overall character of the variability is not constant over the entire
data set - the behavior of the optical light curve in 1990 - 2000 appears 
different from that in 1910 - 1985. The evolution up to 1985 can be 
represented by a single parabolic-type curve with maximum flux of 
about 75 - 80 mJy in B band 
and with superimposed on it more rapid variations.  
In 1935 there was (if the photographic magnitudes were estimated 
correctly) the strongest flare which lasted only 10 - 15 days.  A less
intense flare happened in 1946 but it is based on a single point.  The
evolution in 1990 - 2000 appears as a single giant outburst but
this change in the character of variability occurred in the middle 
of the epoch 1968 - 2000 covered by the photoelectric data.

\subsection{X-ray data}\label{s22}

The data in the 2 - 10 keV X-ray band also came from different 
papers and databases which were collected using different 
satellites. Fluxes were corrected for absorption. 
The observed time coverage is from October 1974. 

The data points during 1975 - 1992 were taken from Papadakis \& 
McHardy (1995). Of their 133 data points, a majority 
came from the Ariel V Sky Survey, and others were taken from 
literature including observations by EXOSAT, GINGA, OSO-8, 
HEAO-1, TENMA and Ariel VI satellites. Some additional points (in 
period May 1987 - May 1995) were taken from Yaqoob \& Warwick
(1991) and Yaqoob et al. (1993) and were also taken from the 
Tartarus Database of ASCA observations 
\footnote{http://tartarus.gsfc.nasa.gov/}. We corrected their 
fluxes for absorption using the HEASARC's online W3PIMMS Version 
3.1 flux converter \footnote{http://heasarc.gsfc.nasa.gov/} as 
well as papers of Weaver et al. (1994) and Edelson et al. (1996). 
Data from the first three years of RXTE observations during 
October 1998 -- November 1999 were taken from Markowitz \& 
Edelson (2001). The observational data given in that paper in 
$counts/s$ were transformed into unabsorbed fluxes via the mean 
luminosity of $300$-day window given in the paper. We supplemented 
those gathered data with three long sequences. Two EXOSAT light curves
were obtained from Ian M. McHardy for the paper by 
Czerny \& Lehto (1997). These data cover 10 and 11 July 1983 and 
15 May 1985, with rate sampling every $100$ s. Third sequence, from 
the ASCA satellite, covers the period from 12 till 23 May 2000, 
and is nearly continuous (except for Earth occultations and the 
passages through the South Atlantic Anomaly).  For that ASCA data 
set, we use the count rate sampling of 32 s.  The measurements 
in $counts/s$ were again converted to fluxes using the mean luminosity 
for an appropriate data sequence.

\begin{figure} 
\epsfxsize = 90 mm 
%\epsfbox{fig1_tot_nufnu_b.eps} 
%\epsfbox{U-XRAY.PS}
%\epsfbox{/home/bcz/4151opt/od_val_18/new-lcux.ps}
\epsfbox{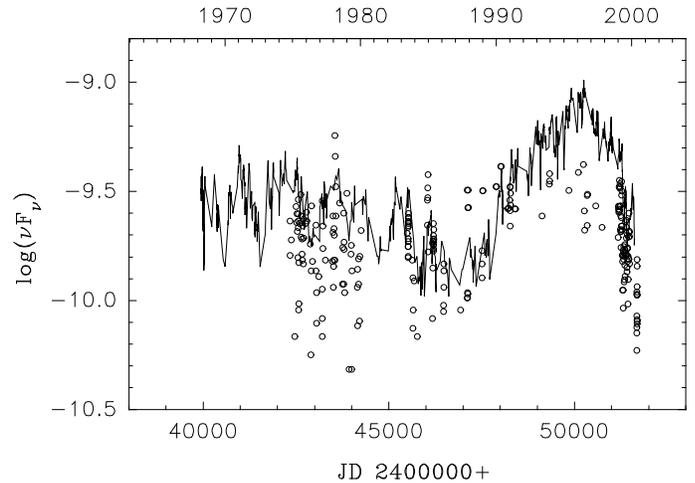}
\caption{The light curve of NGC~4151 ($\nu F_{\nu}$) in U band 
(solid line) and in X-ray band (2 - 10 keV absorption-corrected flux, 
open circles). The errors of X-ray data are on the order of a factor 
2 in the oldest data points and of $\sim 10$ \% for more recent 
measurements. The errors in Crimean data are $\sim 2.2$ \%.  
\label{fig:UXcurve}} 
\end{figure}

All daily-averaged X-ray points used by us for further analysis are 
shown in Fig.~{\ref{fig:UXcurve}} with open circles. 

%UV - any use ??? 
%X-ray - any use (see Warwick R.S. et al. 1996, ApJ, 470, 349) 

\section{Methods}\label{s3}

\subsection{Power spectrum}
\label{sect:metpower}

We attempt the restoration of the underlying power spectrum $|F|^2$ 
of the active nucleus from its observed light curves, i.e.  
from the product of the true light curve and the sampling 
(window) function.  Direct analysis of observations yields 
the observed power spectrum $|G|^2$, i.e.  the underlying 
spectrum affected by the window function $|W|^2$. These functions 
obey an exact relation in the Fourier transform space:
\begin{equation}
   G = F*W 
\label{alex.1}
\end{equation}
where $*$ denotes the convolution and $F$, $G$ and $W$ are Fourier
transforms of the underlying and observed light curves and of the
sampling function. For long and nearly uniformly sampled observations
$|W|^2$ approaches the delta function and $G$ approximates 
$F$. For non-uniform sampling, the shape of the window function becomes
complex, with broad wings and many local maxima. In such case it is
difficult to obtain a solution of Eq. (\ref{alex.1}) for $F$ as it 
involves deconvolution of the sampling transform $W$.  
%In our case this
%is particularly true for the X-ray data, as 2/3 of observations
%coming from EXOSAT are concentrated in two runs of a day length and
%the remaining observations are scattered over several tens of years,
%yielding the X-ray window function with particularly bad
%properties. The former data dominate the high frequency spectrum while
%the low frequency spectrum is dominated by the scattered
%observations.
%To cope with such a situation we apply the code {\sc clean} of Roberts, 
%Leh\' ar \&  Drehner (1987) which determines the true power spectrum
%through the cleaning process based on the 
%Fourier transform of the window function.

We address the issue of uneven sampling by not calculating the power 
spectrum of the entire light curve directly. Instead, we analyze the 
properties in various time scale ranges separately.

In the optical band, we first rebin the data to 1 yr bins and fill the few
existing gaps by linear interpolation, thus obtaining 90 data points. Next
we rebin the original curve into 5 day bins and find an almost continuous
sequence containing 2403 such points. Data gaps are again filled
through interpolation.  We calculate the power spectra of the two sequences 
separately and combine them by averaging the logarithmic values.

In the X-ray band, we also first rebin the data to 1 yr bins (25 points).
Next we rebin the original light curve into 30 day bins and find and almost
continuous fragment (26 points). Next we rebin the original data into 5 day
bins obtaining 69 points in an almost continuous set. Finally, the two 
light curves from EXOSAT and one from the ASCA monitoring were rebinned 
to 4096 s bins. The power spectra of the resulting six light curves 
were computed independently and combined for further analysis.  

For uniformity, all optical data points are converted from magnitudes 
to flux units, when necessary, and the average value is subtracted.
The NPSD of each light curve is computed according to the description by
Hayashida et al. (1998):  
\begin{equation} 
NPSD(f) = {1\over \overline{x(t)}^2} a(f)a^*(f) \times T, 
\end{equation} 
with 
\begin{equation} 
a(f) = {1 \over n} \sum_{i=1}^{n} {x(t_i) \exp(2 \pi i f t_i)}, 
\end{equation} 
where $x(t_i)$ is the value of the flux at the time $t_i$, $T$ is the length 
of the data set and $ \overline {x(t)} $ is the mean value of the flux. 
Therefore, the integrated NPSD gives half of the fractional variance.

The errors of the power spectra obtained as above are either estimated 
directly from the data or estimated using the Monte Carlo simulations 
(see Appendix A).
 
% 7777777777777777777777777777777777777777777777777777777777777777777777777

\begin{table*}     
\caption{The basic parameters of the variability.     
\label{tab:rms}}     
\begin{tabular}{|lcccccc}     
\hline\hline
Color  &  band   &  Flux (mJy)  & rms (mJy) &  $F_{var}$ & $R_{max}$  &  error \\     
\hline\hline
\multicolumn{7}{l}{Whole data}\\ 
\multicolumn{7}{l}{U,B,V (1968-2000), R (1989-2000), X-ray (1975-2001), Btot (1910-2000)}\\
\\
%X-ray  & 2-10 keV& 7.56e-3&3.17e-3& 0.41  & 37.4  & 7.0\% \\ 
X-ray  & [2-10] keV& 2.55e-2$^*$&1.07e-2$^*$& 0.41  & 37.4  & 10.0\% \\ 
U      & 3600 \AA& 48.44  & 24.25 & 0.50  &  8.7  & 2.5\% \\     
B      & 4400 \AA& 63.56  & 21.09 & 0.33  &  3.9  & 1.5\%  \\
Btot   & 4400 \AA& 63.23  & 19.17 & 0.32  &  5.2  & \\        
V      & 5500 \AA& 92.26  & 19.45 & 0.21  &  2.6  & 1.0\%  \\
R      & 7000 \AA& 169.84 & 30.78 & 0.18  &  2.5  & 2.0\%  \\
\hline          
\multicolumn{7}{l}{Cycle $\cal A$ (25.03.1968 - 03.07.1988)} \\
\\                 
U      & 3600 \AA & 31.88 & 10.66 & 0.33  & &  2.5 \%  \\     
B      & 4400 \AA & 48.26 & 9.10  & 0.18  &&   1.5 \% \\     
V      & 5500 \AA & 77.82 & 8.37  & 0.11  &&  1.0 \%  \\     
\hline     
\multicolumn{7}{l}{Cycle $\cal B$ (11.02.1989 - 13.03.2000)}\\
\\    
U      & 3600 \AA & 63.51 &  23.32  & 0.37 && 1.3 \% \\     
B      & 4400 \AA & 75.35 &  20.09  & 0.27 && 0.74\%  \\     
V      & 5500 \AA & 103.43 & 18.18  & 0.18 && 0.83 \% \\           
\hline     
\end{tabular} \\
$F_{var}$ - rms to mean ratio, $R_{max}$ - maximum to minimum flux ratio \\
$^*$ Flux at 2 keV computed assuming $\alpha = 1$ energy index in 2 - 10 keV
band    
\end{table*}     

%7777777777777777777777777777777777777777777777777777777777777777777777777

\subsection{Structure function}\label{s33}

For the analysis of our time series we also used the structure function
technique (SF). The application of SF allows us to study both stationary 
and nonstationary processes. The SF was introduced by Kolmogorov in
1941 for analysis of the statistical problems connected with
turbulence theory (Kolmogorov 1941a;  b).  In astronomical 
practice, the wide application of the SF to time-series analysis
begun in the middle eighties (e.g. Simonetti, Cordes, \& Heeschen
1985; Paltani, Courvoisier, \& Walter 1998; Kataoka et al. 2001).

As a rule, most work in astronomy uses only the first-order structure
function ($SF_1$). By definition the structure function of the
first-order is a mean square of difference $[x(t)-x(t+\tau )]$:
\begin{equation}
 SF_1(\tau) = M[(x(t)- x(t+\tau)]^2,
\label{eq:sf} 
\end{equation}
where $x(t)$ is random process and $\tau$ is time shift.  Since we do not 
use higher order structure functions, we simply denote $SF_1$ as $SF$.  

In the case of a stationary random process the $SF$ is directly related
to the autocorrelation function:
\begin{equation}
 SF(\tau) = 2D[x(t)] \times [1-ACF(\tau)],
\end{equation}
where D[x(t)] is variance of the process and ACF is the
autocorrelation function.

The slope of the $SF$ changes with time interval $\tau$. If the
measurement errors are neglected, on the shortest time scale the
variability can be well approximated by a linear trend, and then
$SF \propto \tau^2$.  For long time scales, the slope of the $SF$
becomes flatter and in the limit when $\tau \rightarrow\infty$ the
structure function saturates: $SF \rightarrow 2D[x(t)]$.  The
addition of measurement errors to the random process increases $SF$
by the value $2D_{err}$ where $D_{err}$ is the variance of the
measurement errors. Therefore, 
$SF \rightarrow 2D_{err}$ when $\tau \rightarrow 0$.  

Many processes are stochastic in nature.  For some processes such as 
fractional Brownian motion, the $SF$ shows a power law dependence
on $\tau$.  Some processes can be represented as a superposition of
a large number of random impulses of deterministic form (shot noise).
Sub-group of those processes which are characterized by the $PSD(f)
\propto 1/f^{\gamma}$  are called flicker noise. If $\gamma$ takes 
values from 1 to 3, then in this case the structure function 
$SF(\tau) \propto \tau^b$ where $b$ takes values from 0 to 2, 
following the relation
\begin{equation} 
\gamma = b + 1. 
\end{equation} 
%This relation does not apply to power spectra flatter than $\gamma = 1$. 
%A case of $\gamma = 1$ is already degenerate in a sense that the
%corresponding $SF$ is flat, $(b=0)$, and the white noise signal
%($\gamma = 0$) also leads to the flat SF $(b=0)$.

The structure function analysis was done using a software package by
S.G. Sergeev from the Crimean Astrophysical Observatory. The whole time
interval was divided into equal bins in logarithmic scale and for each 
bin we found such pairs of observations with $t_j > t_i$ that their 
time difference $\tau_k = t_j - t_i$ fitted into the given bin. Next 
we calculated the value 
\begin{equation} 
SF (\tau_k)=\sum_{i,j} {[x(t_j) - x(t_i)]^2 \over n_k}, 
\end{equation} 
where $n_k$ is the number of pairs in k-th bin.  
We estimate the errors of the SF as described in the Appendix B.

To allow a meaningful comparison of the SF in various bands we 
introduce also the normalized structure function (NSF) as follows:  
\begin{equation}
NSF(\tau) = {SF(\tau)\over D[x(t)]}.
\label{eq:NSF}
\end{equation}
In this case all NSF should approach the universal value 2 on long timescales.

\subsection{Autocorrelation function}\label{s34}
 
%Generally, for computing the autocorrelation function (ACF) or the
%cross-correlation function (CCF),  either interpolation or 
%discrete methods are used (see Edelson \& Krolik 1988). 
%Interpolation method is based on the linear 
%interpolation of a light curve to obtain its value after the shift
%by the time 
%interval $\tau$, and discrete method is based on collecting the pairs of 
%points corresponding to given bin of time interval $\tau$. 
%Discrete method allows to calculate the errors for correlation 
%coefficient but interpolation method is more effective.
In our computation of ACF, we again  use of program package by 
S.G. Sergeev;  the method used relies on the interpolation.  
Specifically, for any point from the real data, a corresponding
shifted point is found by linear interpolation or extrapolation,
and extrapolation is made adapting 
$x(t>t_{max})=x(t_{max})$ and $x(t<t_{min})=x(t_{min})$ (for a more
detailed description, see Sergeev et al. 1999). 

\section{Results}\label{s4}

\subsection{General variability properties of NGC~4151}\label{sect:gen}

The basic variability properties of the Seyfert galaxy NGC~4151 seen in the
data assembled by us are summarized in the upper part of the 
Table~\ref{tab:rms}.  We give there the mean flux at the given 
spectral band, rms fluctuations, the normalized amplitude 
$F_{var}$ (i.e. rms to mean ratio) and the ratio $R_{\max}$
of the maximum to the minimum flux. The full 90 years of the B 
band data are given separately from the data subset determined 
by photoelectric measurements.

The normalized variability amplitude is the largest in U band and decreases
towards the longer wavelengths. This effect is well established for AGN
(e.g. Ulrich et al. 1997) and tells us that the variable component is
bluer than the non-variable (starlight) component.  

The overall normalized variability amplitude in the X-ray band is comparable
to the amplitude in U band. However, more extreme variability events
occasionally happen in X-ray band while the systematic long timescale 
dimming/brightening seems to be stronger in U band than in X-rays. 
The energy content is significantly greater in the variable part of the 
optical spectrum than in the X-ray spectrum. Variable part of 
$\nu F_{\nu}$ in U band is equal to $2.02 \times 10^{-10}$  erg 
cm$^{-2}$ s$^{-1}$ while the variable part of the 2 - 10 keV flux is only
$0.81 \times 10^{-10}$ erg s$^{-1}$ cm$^{-2}$. 
Of course we do not know the bolometric corrections needed for more 
strict comparison of the total X-ray and optical variable energy 
output but, they are not likely to reduce this discrepancy considerably. 
Therefore, we conclude that strong long timescale trends in the optical 
band do not result 
from reprocessing of the variable X-ray emission, 
unless X-ray variability leads to strong variations of the extinction, 
thus amplifying the effect.

\subsection{Stationarity of the data}\label{sect:stat}

Question of whether light curves of AGN covering a time span of 
several decades may be considered realizations of a stationary process is 
interesting {\em per se}, but also because of its possible influence on
investigations of variability on shorter time scales 
(e.g. Leighly 1999, Press \& Rybicki 1997). 
We address this issue by estimating of the importance of any existing linear
trend.

\subsubsection{Optical band}\label{s411}

Even the visual inspection of the light curves of 
NGC~4151 clearly shows some long timescale trends 
although they become less prominent with an increase of 
the data length. It can be best studied in the B band 
since at that color the available light curve covers 
the longest time span.  

We analyzed the longest trends by rebinning the data in one year 
bins in order not to be biased by the most recent time spans with 
superior data coverage as compared with the earliest data 
from the beginning of the previous century. We studied the trends both 
using magnitudes (effectively logarithmic scale) or using $\nu 
F_{\nu}$ fluxes (effectively linear scale). 

Mean value of the B magnitude in the Crimean observations is 11.83. 
Apart from two major outbursts, the presence of the linear 
trend is also seen in the data. Linear fit in magnitude scale 
gives a shift by -0.35 mag during the entire period of 
observations (32 years). This is comparable to the value of the 
dispersion in this period, calculated for unbinned data (0.33 mag). 
The result is similar if flux ($\nu F_{\nu}$) is used: linear 
trend gives the net brightening by $1.2 \times 10^{-10}$ erg 
s$^{-1}$ cm$^{-2}$, with the dispersion equal $1.4 \times 
10^{-10}$ erg s$^{-1}$ cm$^{-2}$. However, if the entire data set 
of 90 years of data is analyzed, the linear trend acts in the same 
direction but is much weaker. Systematic shift is only -0.13 mag, 
or $3.6 \times 10^{-11}$ erg s$^{-1}$ cm$^{-2}$, with the 
dispersion unchanged. 

This shows that the data are not strictly stationary even on the 
timescale of 90 years. However, in the longest data set the linear 
trend is much weaker than the typical amplitude of the variability 
(in B band) and the criterion for stationarity is roughly satisfied.
It is less so in the case of shorter sequences, so the power spectra 
and the structure function may depend on the choice of 
the data set. This is particularly well known from the studies of X-ray 
light curves of galactic sources (e.g. Uttley \& McHardy 2001, and 
references therein; for quasar sample see Manners, Almaini, 
\& Lawrence 2001). 

\subsubsection{X-ray band}\label{sect:statX}

X-ray data cover only the period from 1974 till 2000, and the 
older data (till $~\sim 1986$) seem to show larger scatter so the 
presence or the absence of the long trends is less apparent (see 
Fig.~\ref{fig:UXcurve}).  
 
We again rebinned the data in one year bins in order to achieve 
a uniform distribution on long time scales and looked for a linear 
trend in the resulting light curve. The mean 2 - 10 keV flux was 
$1.98 \times 10^{-10}$ erg s$^{-1}$ cm$^{-2}$ while the 
linear trend gave the systematic brightening of the source by 
$4.12 \times 10^{-12}$ erg s$^{-1}$ cm$^{-2}$, much lower than the 
dispersion in the data during this period ($8.1 \times 10^{-11}$ 
erg s$^{-1}$ cm$^{-2}$).  Such absence of the linear trend in the 
data is partially accidental - more accurate Ginga/ASCA/RXTE data 
(1978 - 2000) cover the period which started and ended up at a very
similar level of the activity of the source. Slightly 
shorter sequences show trends up to almost an order of magnitude 
stronger but still smaller than the dispersion. 

This suggests that the X-ray flux is more likely to be produced by a 
stationary process during the observed period than the optical flux 
during the same epoch. 
%*** For a stationary time series PSD must saturate at sufficiently long time scales (c.f. Eq. 9). 
Note however, that the time scales used for sampling of the X-ray flux 
range from seconds up to decades, i.e. have a greater dynamical range than the 
time scales of optical sampling, which range from days up to nearly a 
century.  

In this context a detection of the change of slope (and in particular, 
of flattening) of the PSD towards longer time scales might be easier 
in X-rays than it would be in the optical band.  A few more
years of occasional monitoring are needed to confirm such a statement. 

\begin{figure} 
\epsfxsize = 90 mm 
%\epsfbox{histc.eps} 
%\epsfbox{power2.eps}
\epsfbox{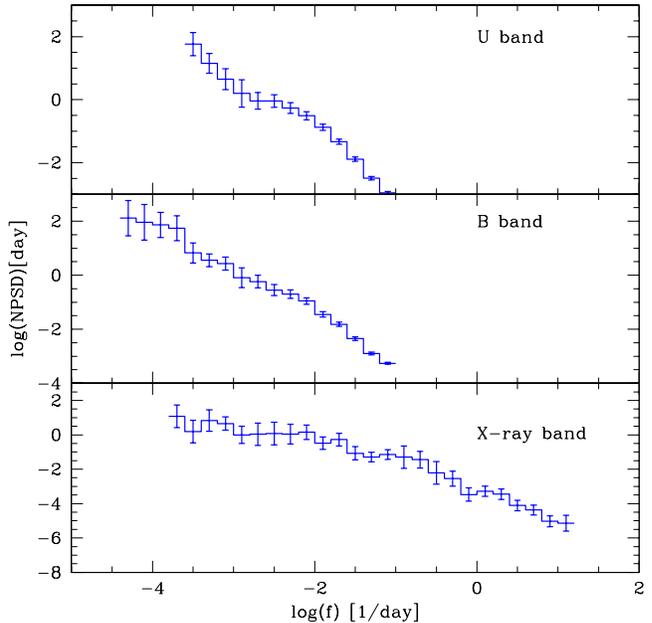} 
\caption{Normalized Power Spectrum Density of NGC~4151. 
Upper panel: U band (1968 - 2000); middle panel:  B band (1910 - 2000); 
lower panel: X-ray band (1974 - 2000). Marked errors are direct 
observational errors, as described in Appendix A.  
\label{fig:power}} 
\end{figure} 

\begin{figure} 
\epsfxsize = 90 mm 
\epsfbox{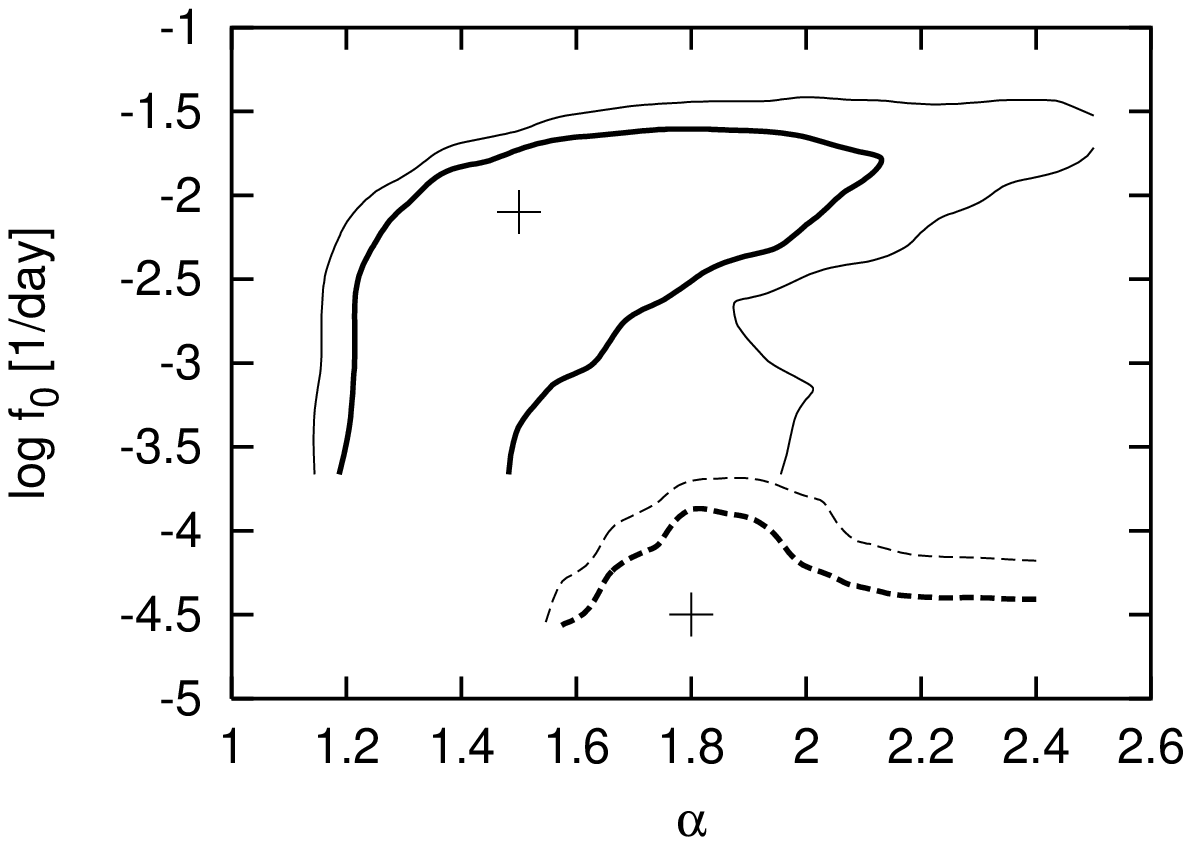}
\vskip -2 truecm
\epsfxsize = 90 mm
\epsfbox{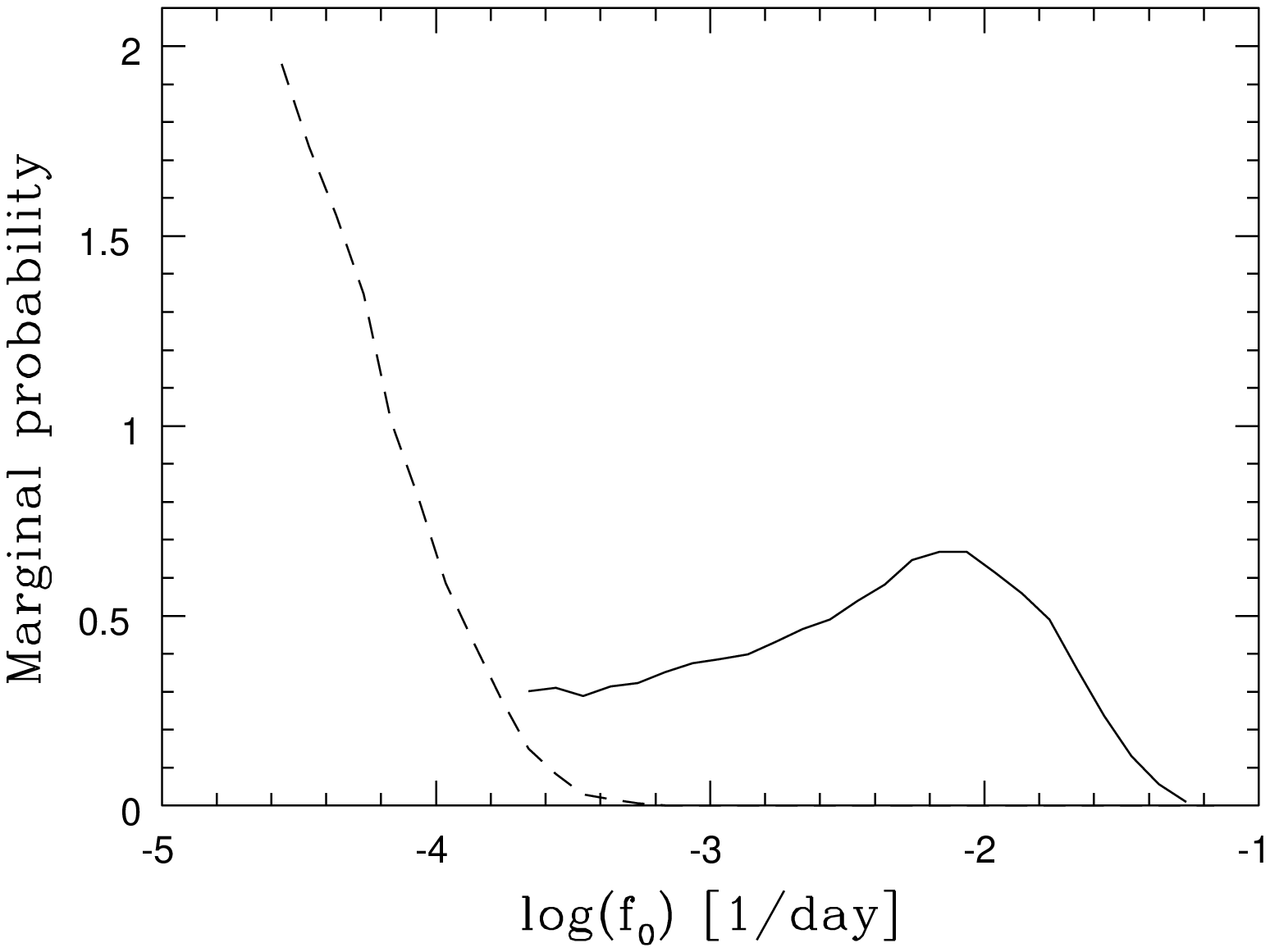}
\vskip -3.5 truecm
\epsfxsize = 90 mm
\epsfbox{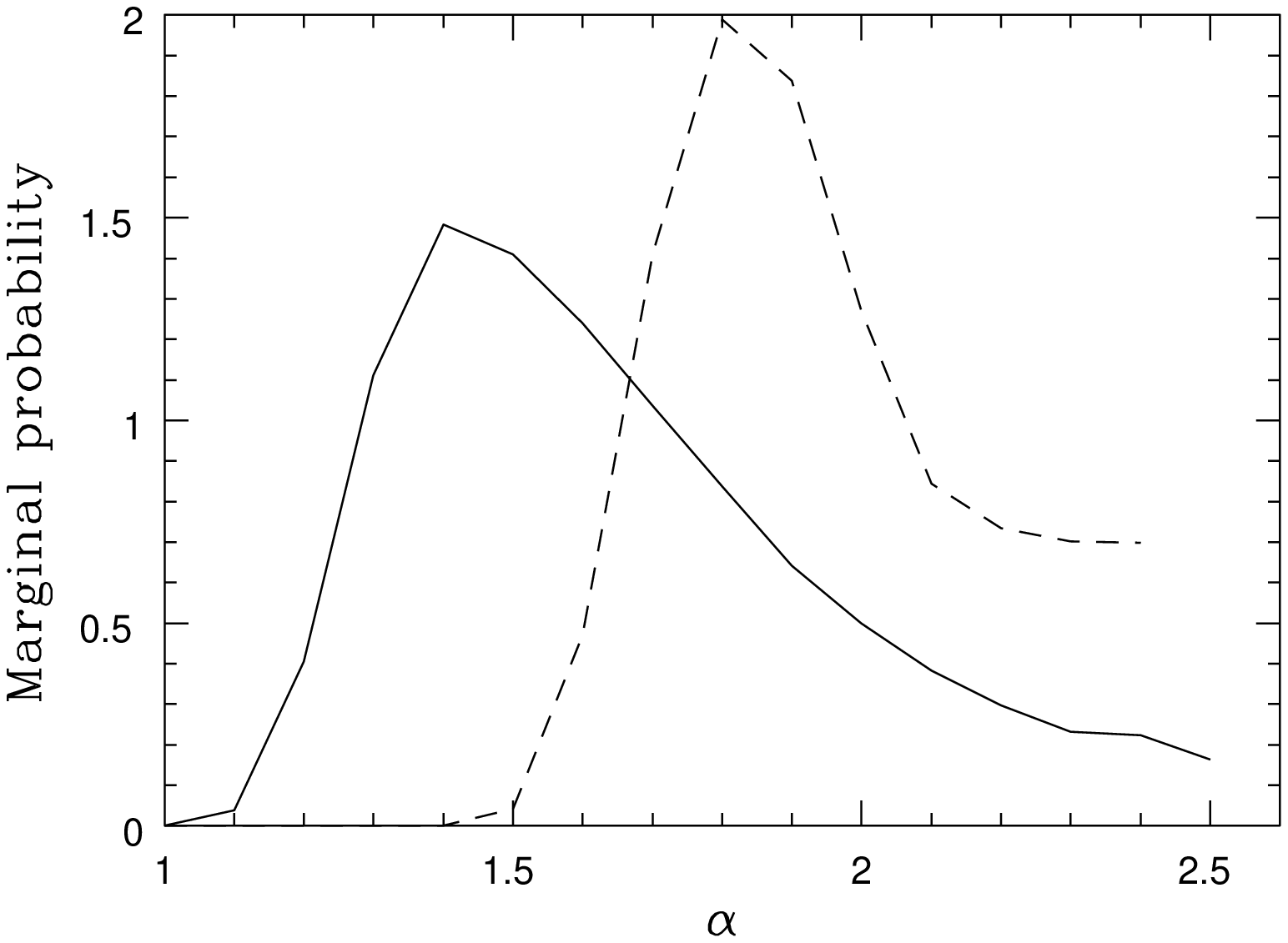}
\vskip -1.5 truecm 
\caption{Upper panel: error contours (thick line - 68\% confidence level,
thin line - 90\% confidence level) for the fit of the power law with an
index $\alpha$ and a break at $\log f_0$ to the NPSD of NGC~4151
in the B band (1910-2000; dashed line) and X-ray band (continuous line) determined 
from Monte Carlo simulations (see Appendix A). Cross marks the location 
of the best fit.  Intermediate and lower panels show the 
plots of the marginal probability for both parameters in the X-ray 
band (continuous line) and in the optical band (dashed line).  
\label{fig:err1}} 
\end{figure} 

\subsection{Power spectrum}\label{s42}

\subsubsection{Optical band}\label{s421} 

We analyze first the Crimean data alone, covering the years from 1968 
till 2000, in all three color bands. 

The power spectrum in the U band (see Fig.~\ref{fig:power}, top panel)
has roughly a power law shape with a steep slope ($\sim 2$) but the 
presence of some structure is clearly visible. A flattening is seen 
at a timescale of $\sim 300 $ d which seems to separate two variability 
components. No flattening is seen at the longest timescales but the 
data set used to determine this power spectrum covers only $\sim 10$ 
years.  In the V band, the spectrum (not shown) shows practically 
no substructure.  Given all this, we infer that the interesting 
timescales are rather long, and therefore concentrate on the results 
derived from the complete set of data in B band, covering the entire 
90 years.  The result is shown in Fig.~\ref{fig:power}, middle panel.
In similarity to the U band power spectrum, the B band data 
show a hint of the presence of two components, separated at the 
timescale on the order of $\sim 300 $ days. The data  
also reveals the flattening of the low frequency component 
(roughly at timescale of $\sim 10$ years) although the measurement 
errors are large.  

In order to determine whether the flattening of the power spectrum 
is real and not simply caused by the finite extension of the data, 
we performed Monte Carlo simulations assuming that the power 
spectrum of the entire 90 years of the data in B band is represented 
by a broken power law model
\begin{equation}
 f \propto \left\{\begin{array}{ccc} const ~~~ & {\rm if } ~~~ \log f <f_0\\

           f^{-\alpha}       &  {\rm if } ~~~ \log f > f_0\end{array}\right.
\label{eq:broken}
\end{equation}

The best fit values of the model parameters are given in 
Table~\ref{tab:psfit}, together with 1 $\sigma$ errors for 
each parameter. Full two-dimensional contour errors are shown 
in Fig.~\ref{fig:err1}.  Despite the fact that the data used for this 
cover such an exceptionally long period, the errors are still 
considerable:  the NPSD is consistent with a single power law. 
It is steeper than 1.5 and the flattening happens at timescales 
longer than 15 years. We note here that the outlined 
procedure for computation of the confidence levels is based on likelihood 
function derived from simulations. Conceptually it is different from the 
$\chi^2$ method employed in Appendix A. The two methods converge only in 
an asymptotic limit of infinite number of observations. Here we adopt the 
former method as the more conservative one, yielding much broader confidence 
limits. 

%88888888888888888888888888888888888888888
\begin{table*}
\caption{The NPSD parameters determined from Monte Carlo simulations for 
a single break model of 1910-2000 B light curve and two 
models of the X-ray light 
curve, one with a single and another with a double break
\label{tab:psfit}}     
\begin{tabular}{lcccccc}
\hline\hline
Band  &  slope   &&&&& $\chi^2$ \\
\hline\hline
Btot     & $ 1.8^{+0.3}_{-0.1} $ &&&&& 5.14/15 dof \\
X-ray     & $ 1.3^{+0.5}_{-0.1} $ &&&&& 20.16/23 dof \\
\hline\hline
Band  &  slope   & $\log f_0 [d^{-1}]$   &  slope &&& $\chi^2$ \\
\hline\hline
Btot     &   0      &$ -4.3^{+0.8}_{-\infty} $ & $ 1.8^{+\infty}_{-0.3}$ &&& 5.14/14 dof\\
X-ray     &   0      &$ -2.1^{+0.5}_{-1.2}   $  & $ 1.5^{+0.5}_{-0.2}   $ &&& 12.39/22 dof\\
%\end{tabular}
%\begin{tabular}{lcccccc}
\hline\hline
Band  &  slope   & $\log f_1 [d^{-1}]$   &  slope  &  $\log f_2 [d^{-1}]$  & slope & 
$\chi^2$ \\
\hline\hline
X     &   0      &$ -3.1^{+1.1}_{-\infty} $&   1  & $ -0.7^{+0.6}_{-0.7}$ & 2 & 20.31/22 dof \\
\hline     
\end{tabular} \\
Quoted errors are determined from $\chi^2_{dist}$ statistics (see Appendix A)
and
correspond to 1 $\sigma $ error for one parameter of interest.
\end{table*}
%8888888888888888888888888

\begin{figure} 
\epsfxsize = 90 mm
%\epsfbox{figtotfreq.eps}
\epsfbox{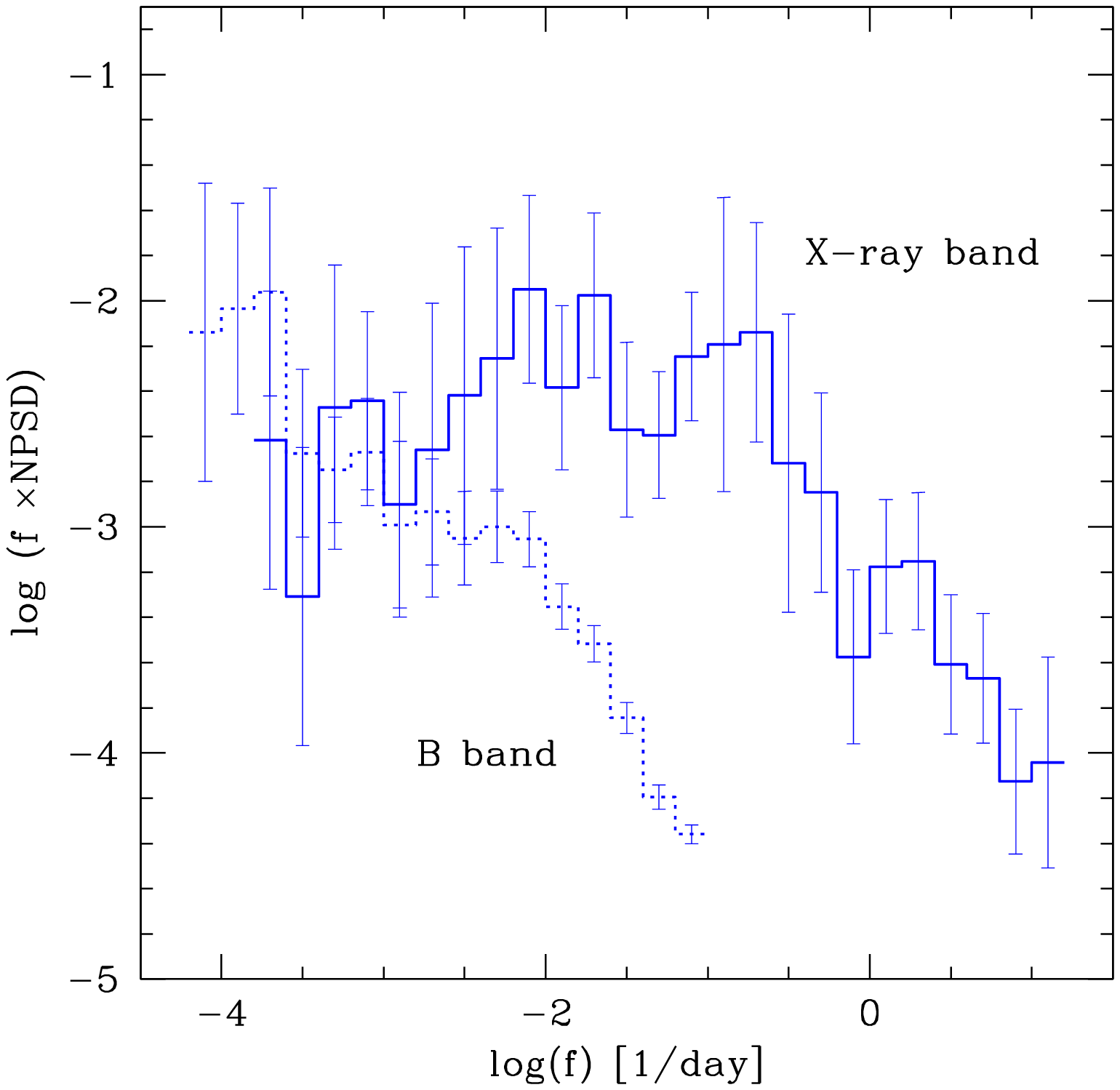}
\epsfxsize = 90 mm
\vskip  -1 truecm
\epsfbox{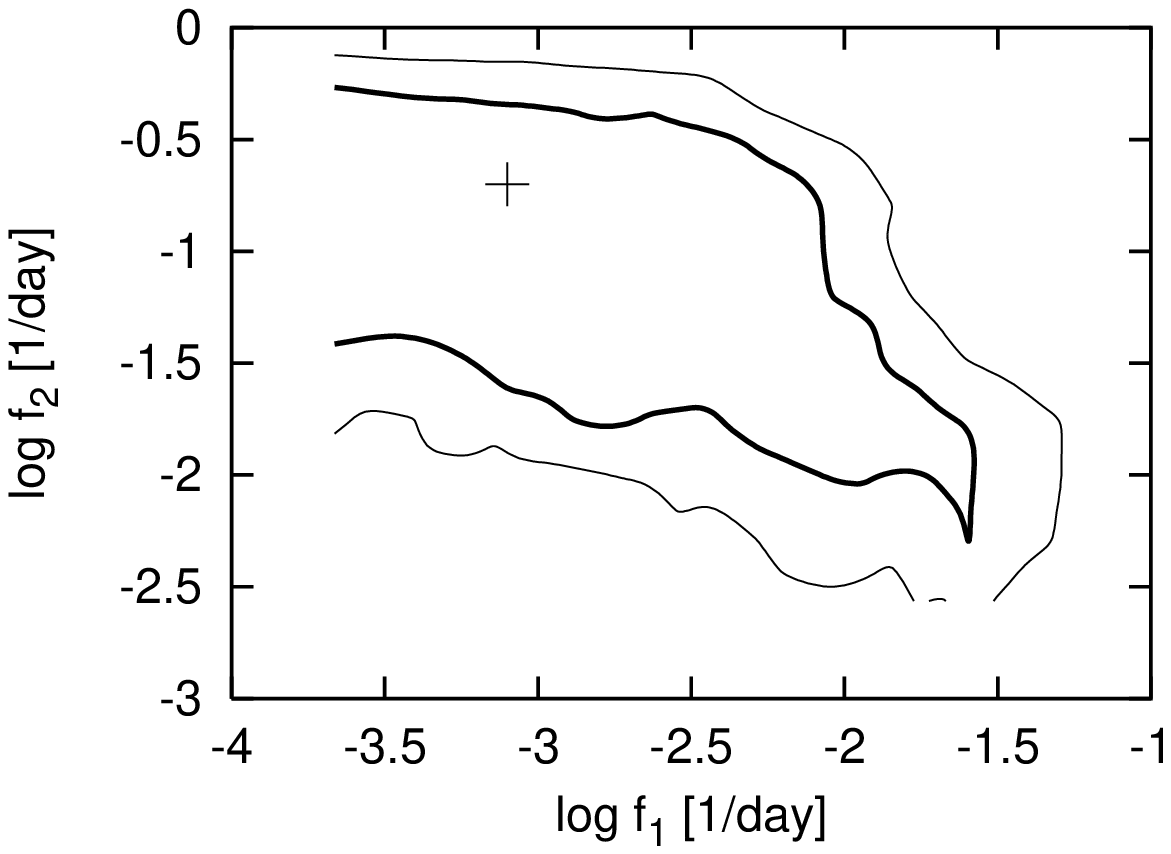} 
\caption{A plot of NPSD multiplied by the frequency, illustrating 
the frequency regime where most of the power is present: dotted line 
-  B band (1910 - 2000), continuous line - 2 - 10 keV band. 
Lower panel shows the error contours for the fit of the
NPSD with two broken power laws of slopes 0, 1 and 2, and the break
frequencies given by $\log f_1$ and  $\log f_2$. Cross marks the best 
fit position.  
\label{fig:nupower}} 
\end{figure}

We reploted the resulting NPSD in the form of $Power \times Frequency$
diagram (see Fig.~\ref{fig:nupower}) in order to better illustrate 
the dominant time scales.  The component corresponding to the timescale 
of years appears to dominate, but the Figure also shows the presence of a 
separate, month-timescale component, as well as a possible high frequency
roll-over to the last component at about $\log f$(days)$ =-2$. It would agree
with the characteristic timescale determined by Collier \& Peterson (2001).

Lyuty \& Doroshenko (1999), analyzing the data from the period 
1968 - 2000, argued that there were two cycles of activity in NGC~4151 
which differed considerably regarding the level of variability: first 
one from 1968 till 1988 (cycle $\cal A$) and other from 1989 till the 
present time (cycle $\cal B$).  Since the duration of the outburst 
$\cal B$ - the dominant feature in the optical light curve (see 
Fig.~\ref{fig:Bcurve}) - corresponds to the characteristic frequency 
seen in our analysis, we repeated this analysis for the data set in 
the B band without the cycle $\cal B$.  The NPSD computed without the 
cycle $\cal B$ still showed a flattening at $\log f$(days)$ \sim -3.6$ 
but above it the NPSD was even better represented by a single power 
law with a slope $\sim 2$.

%\begin{figure} 
%\epsfxsize = 80 mm \epsfbox[50 180 560 660]{computations/Fig9.eps} 
%\epsfxsize = 90 mm
%\epsfbox{figtotfreq.eps}
%\epsfbox{clean/test2.eps} 
%\caption{The NPSD multiplied by the frequency showing comparison
%between the results with cycle B data and without it. 
%\label{fig:bezB}} 
%\end{figure} 
  
\subsubsection{X-ray band}\label{s422}
 
X-ray PSD is distinctively different from the optical NPSD (see 
Fig.~\ref{fig:power}). Most of the power is now on much shorter 
timescales. The spectrum generally shows a flattening on time scales of 
about 100 days, but more structure can be identified.  The spectrum 
might be crudely represented as a broken power law, with slope 1 at 
intermediate frequencies and about 2 at higher frequencies.  The 
normalization at high frequencies is similar to that obtained by 
Hayashida et al. (1998).  The position of the flattening determined by
Papadakis \& McHardy (1995) is now in the middle of the intermediate part. 

The characteristic features of the shape are again apparent on 
$Power \times Frequency$ plot (see Fig.~\ref{fig:nupower}). Most of the 
X-ray power is concentrated between $\sim 6$ days and $\sim 150$ days, 
with a turnover both at higher and lower frequencies. There may also be a 
decrease in the variability level at the intermediate timescales of 
$\sim 30$ days. There may also be some additional power at the
lowest frequencies but at the lower level than in the optical band.

Since the effect of the specific window function is even stronger for X-ray
data than for the optical data, we again performed Monte Carlo simulations,
as described in Appendix A. We considered two models for the shape of 
the NPSD.  

The first model was a broken power law identical to the adopted model for the
optical power spectrum (see Eq.~\ref{eq:broken}), with the frequency break 
and
the slope above it being the free parameters of the model. The slope obtained
from simulations (see Table~\ref{tab:psfit}) is slightly flatter than the
optical one, but the difference is within 1 $\sigma $ error. However, the
best fit break frequency of the NPSD in the X-ray band corresponds to
a timescale of around $\sim 100$ days. Formal contour errors slightly overlap
(see Fig.~\ref{fig:err1}), particularly at the 90\% confidence level but the
probability maxima are well separated. The marginal probability 
plots made by 
integrating the probability distribution along one of the parameter axis 
(see the intermediate panel of Fig.~\ref{fig:err1}) show that the 
distributions of the slopes in the optical and X-ray NPSD are only slightly
different, namely they are shifted by $\sim 0.4$ decades, while the 
distributions of the characteristic variability timescale
are essentially different. The value on the order of 100 days is favored for
X-ray data but for the optical data the maximum coincides with the longest
timescale under study ($\sim 100 $ years). It strongly supports the view
that the X-ray and optical variability properties at long timescales are
significantly different. 

The second model was the shape frequently adopted for the galactic sources,
a power law distribution with two breaks but fixed slopes of 0, 1 and 2 at
the lowest, intermediate and the highest frequencies. Such a model is also a
good representation of the observed power spectrum, although formally, 
the best fit model has a lower acceptance probability than in the previous 
case. The break frequencies in the best fit model correspond to the 
timescales of $\sim 6 $ days and $\sim 1000$ days. The low frequency break 
is at lower frequencies than expected from Fig.~\ref{fig:nupower} since 
apparently the extra power at the lowest frequencies shifts it towards 
longer timescales.

The spectrum may perhaps be better represented by a number of Lorentzian
components, as suggested e.g. by Pottschmidt et al. (2002) for galactic 
sources but the quality of our light curves was not high enough 
to attempt more sophisticated fits.

\subsection{Structure function}\label{s43}

The analysis via the structure function was applied to the Crimean 
data covering the years from 1968 till to 2000 in UBVR bands, 
to the combined photographic plus photoelectric data in B band 
covering the years from 1910 till to 2000, to the AGN Watch data, and also 
to the X-ray data collected from the literature. 

\subsubsection{Optical band}\label{opt}

\begin{figure} 
\epsfxsize = 90 mm 
%\epsfbox{N4151SF.PS} 
%\epsfbox{SF-N4151.ps} 
%\epsfbox{Sf-n4151.ps}
\epsfbox{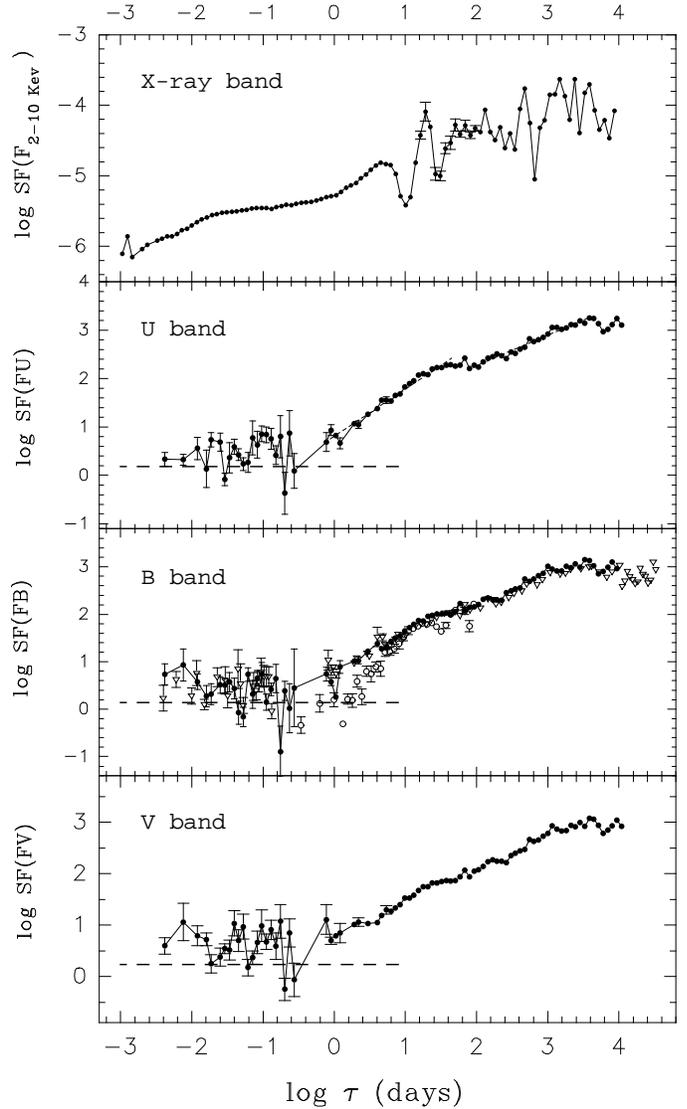} 
\caption{The structure function for NGC~4151: upper panel - X-ray band,
second panel - U band, 
third panel - B band (Crimean data - filled circles; AGN 
Watch - open  circles; photographic data - triangles), 
lowest panel - V band.
Errors given here are statistical errors (see Appendix B).
Dashed line shows the level of SF due to the measurement error in the data.
\label{fig:SF}} 
\end{figure} 
 
Fig.~\ref{fig:SF} shows SFs for the optical flux in various bands, 
where the Crimean data is plotted with filled circles.  One can see that the 
shape of the SF which would have been inferred from the Crimean V magnitude 
data alone can be very well described as a single power law form, with the 
slope $b \approx 0.7$ for time scale of variability from 
1 to 3200 days.  Such value of $b$ shows that the dominating process 
is a single shot noise process. But in the U and B bands, the shape of 
the SF is more complex, and a clear break in the SF is seen at about 
30 - 50 days, with much steeper slope at short timescales and a flatter 
one in long timescales. The results of Merkulova et al. (2001) 
show that the steep slope continues down to the timescales of minutes. 
It can again indicate that in these two time intervals two different 
% shot noise 
processes operate. We therefore measured separately the slope in 
time interval from 1 to 40 days and in time interval from 80 to 
3200 days (see Table~\ref{tab:sfAB}). Again, in the V band, a 
single slope of 0.7 provides a good representation of the 
variability in all timescales shorter than 3200 days (8.8 years) 
but at shorter wavelengths the two slopes are significantly 
different. This difference is most visible in U band. 
The traces of similar structure are only barely seen in V band. 

We treated separately the SF of AGN Watch data.  Continuum 
fluxes at $\lambda$ 4600 \AA\ were converted to fluxes in B band
and plotted as open circles in the third panel of Fig.~\ref{fig:SF}). 
The corresponding points match well the Crimean data above the 
timescales of $\sim 10$ days but they steepen significantly below.  
This effect is probably due to the presence of a strong linear trend in 
the optical data during the period covered by AGN Watch monitoring 
(see Fig.~\ref{fig:Bcurve}). Such a linear trend translates into 
$SF \propto \tau^2$, as can be shown analytically from Eq.~\ref{eq:sf}. 
This kind of discrepancy underlies the importance of using as long time 
series of observations as possible in order to avoid artifacts.  

One of the important characteristics of the structure function is the
point which marks the beginning of plateau at the long time lags. This
time scale gives the maximum time scale T$_{max}$ of correlated 
signals or, equivalently, the minimum time scale of uncorrelated behavior. 
From Fig.~\ref{fig:SF} we can see that the SF reaches plateau 
at $\log\tau\approx 3.2 - 3.6 $, i.e. about 1600 - 4500 days.  
We can estimate it most accurately in the B band, due to the
longest time coverage in this band.
%It is possible this time scale connected with the 
%active phase which continues from 1989. 
Since the photographic data is of relatively low accuracy,
and the number of data points is small in comparison with 
photoelectric measurements, we compute separately the SF only 
for historical data. This structure function 
%corrected for the constant 
%contribution due to measurement error (8.5\%) 
is shown in the third panel of Fig.~\ref{fig:SF} with open triangles. We see 
that the slopes of structure functions derived from both data sets 
are similar within the errors.
%Rather flat SF slopes in 80-4500 days range is present in 
%the long data set.
Although the length of historical data set is about 25200 days 
($\log \tau =4.4$), the plateau begins earlier, at about $\log \tau=3.2$ 
(corresponding to the time interval 1600 days). So the existence of the 
plateau is not an artifact of an insufficiently long data train, 
but its presence is supported by long observations. 

The flattening observed in SF at 1600 - 4500 days (4 - 12 years) 
corresponds well to the flattening hinted by the optical NPSD,
at $\sim 10$ years.  The change of the slope at some temporal frequency 
is also seen in both cases, although it does not happen 
precisely at the same timescales. SF suggests a break at 
$\sim 30$ while that apparent in the NPSD is at $\sim 500$ days.  

Although the SF is less sensitive to the window function problem than the
NPSD technique, we performed some Monte Carlo simulations to estimate
better the uncertainties involved in the analysis.
We adopted a shot noise model of the light curve, with the model parameters
\begin{equation}
dT,\alpha,\beta,T_{min},T_{max}; ~~~~~  b=2 + \alpha + 2 \beta,
\end{equation}
being the mean time separation of the flare, power law index of flare
number
distribution, power law index of flare duration distribution, minimum and
maximum duration of a flare, correspondingly (see Appendix B).

\begin{figure} 
\epsfxsize = 85 mm
\vskip -5 truecm 
\epsfbox{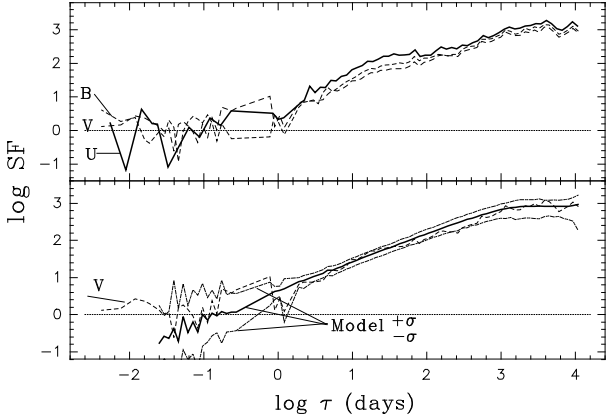} 
\caption{Upper panel: the structure function in three 
color bands for the Crimean data (U - continuous line, B - thick dashed line,
V - thin dashed line), with the contribution from observational errors
subtracted. Lower panel: comparison of the SF for the data (dashed
line) and for the best fit model (thick line) in the V band. The error on the
simulated SF in the lower panel is determined from the dispersion in Monte Carlo simulations (see
Appendix B). Dotted lines show the level of SF due to the measurement
error in the data.
\label{fig:sim3}} 
\end{figure}

The best fit model parameters representing the Crimean data in the 
U, B and V bands are given in Table~\ref{tab:sffit}), and the best fit 
models are shown in Fig.~\ref{fig:sim3}.  There is a good overall 
agreement between the SF for the best fit model and the SF derived 
from the data (see lower panel of Fig.~\ref{fig:sim3} for comparison 
of the observed and modeled SF in V band).  The SFs in all
three colors are basically similar to each other although hints of systematic
differences may be seen at the intermediate timescales between 1 and 60 days,
reflecting the minor difference in the best fit parameters. Those differences
are marginally significant. In the lower panel of Fig.~\ref{fig:sim3} we show
the dispersion of the SF for the family of Monte Carlo curves in the V band;
the indicated error is about 0.2 in logarithmic scale at intermediate 
timescales and becomes larger at long and short timescales.

\begin{figure} 
\epsfxsize = 85 mm 
\epsfbox{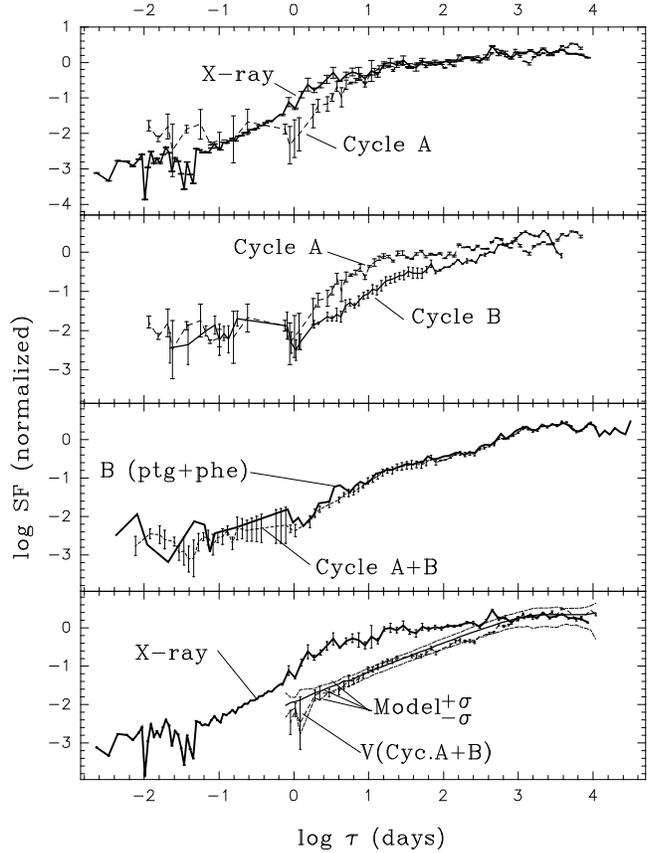} 
\caption{Normalized SF for observational data. 
Upper panel: results for X-ray band with long timescale trends 
determined from daily averages, and short timescale trends from 
the ASCA data alone, together with the SF for the cycle $\cal A$ 
averaged over UBV band. 
Second panel: comparison of SF averaged over UBV bands for the cycles 
$\cal A$ and $\cal B$. 
Third panel: comparison of the result from all combined photographic (ptg) 
and photoelectric (phe) data in B band with photoelectric Crimean B band 
data alone. 
Lowest panel: comparison of the X-ray SF with V band SF 
based on Crimean data (see Fig.~\ref{fig:sim3}) and best fit V band model;
the envelope of the V curve marks the error estimate from Monte Carlo 
simulations.
\label{fig:sim7}} 
\end{figure}

We also used all the data collected in the B band in order to exploit 
fully the data length. The result is shown in Fig.~\ref{fig:sim7}.  
In this Figure we plot the normalized structure function for better 
comparison between various curves, and we also subtract the value 
$2 \sigma_{err}^2$, i.e. twice the variance of the measurement 
uncertainties. On the old photographic data we adopt a measurement error 
of 8.5\%.  We see that the SF is not significantly different from the 
one obtained from the Crimean data alone.  However, the apparent 
flattening of the SF at the longest timescales is still
better visible due to the access of the longest timescales.

\begin{figure} 
\epsfxsize = 90 mm 
%%\epsfbox[100 200 400 560]{sfa.eps} 
%%\epsfbox{SF-CYCA.ps} 
\epsfbox{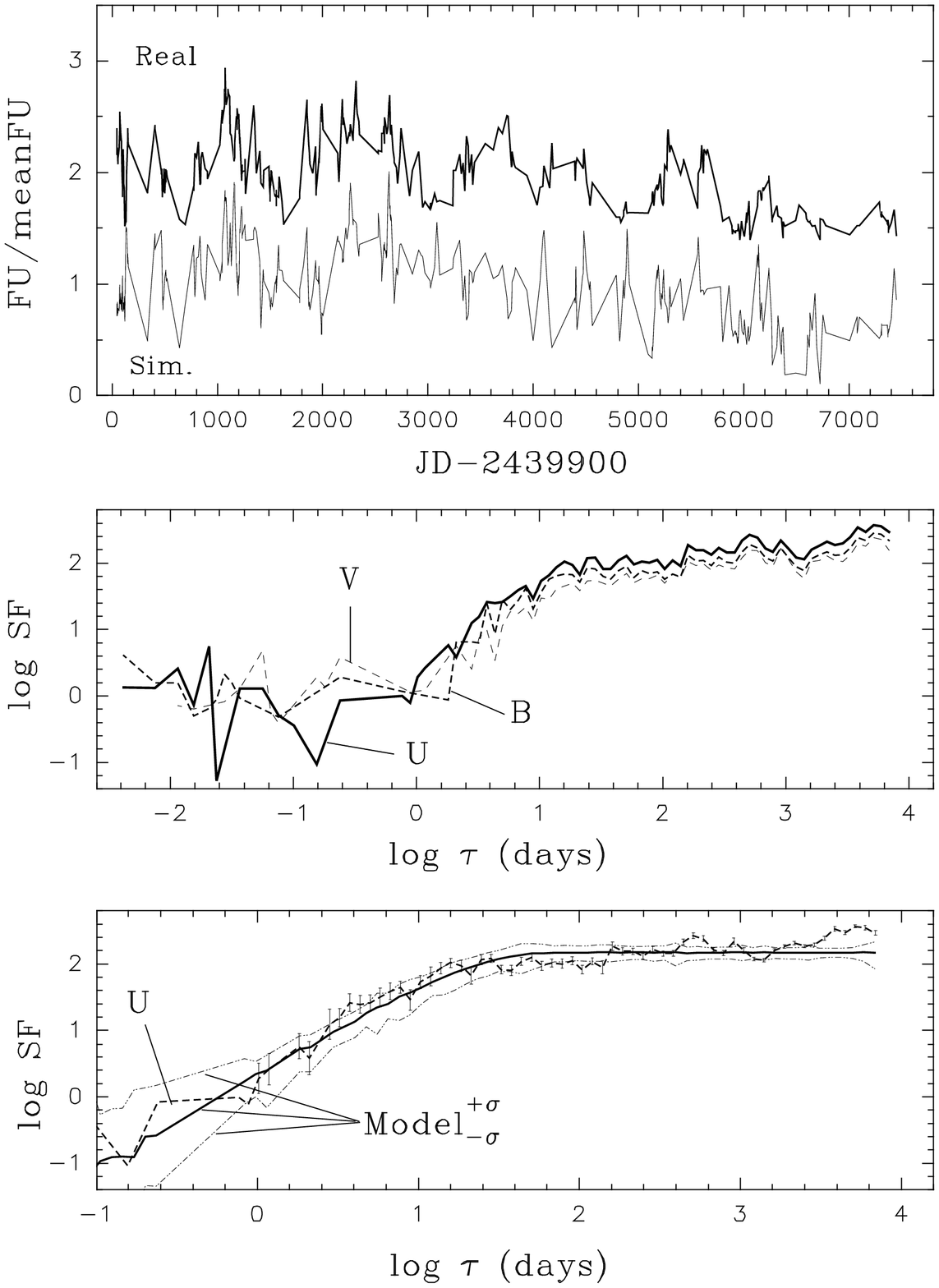} 
\caption{Optical light curve properties during the cycle $\cal A$. 
Upper panel: the observed light curve in U band and a representative light 
curve from Monte Carlo simulations, normalized to the mean flux and 
offset for clarity. Middle panel: observed SF in three color bands. 
Lower panel: mean and dispersion of the SF from Monte Carlo simulations 
for best fit model parameters, together with the observed SF in the U band. 
\label{fig:sfa}} 
\end{figure}

An interesting question is whether the variability properties 
during flares are different, and to address this, we repeat our 
analysis for the two cycles $\cal A$ and $\cal B$ separately. 
All slopes of the SF are presented in Table~\ref{tab:sfAB}. 
An inspection of  Table~\ref{tab:sfAB} reveals that 
the properties of the light curves ($SF$ slopes) are 
significantly different in the two cycles. During the cycle $\cal B$
the slope difference between the short and long timescales is 
moderate, although a feature around $\sim 50$ days is present.
The plateau for long time intervals 
begins at $\log \tau$=3.2 - 3.4. During the cycle $\cal A$, the difference 
of the slopes is apparent in all color bands, with the long timescale 
slope being very flat (Fig.~\ref{fig:sfa}; see also Fig~\ref{fig:sim7},
second panel), suggesting that long timescale trends are relatively 
weak. Monte Carlo simulations with short lasting shots (20
days and less) reproduced well the observed light curve properties.
This is caused by the fact that in the cycle $\cal A$ we do 
not see a single large outburst but instead, a number of weaker 
eruptions superimposed on a weak longer trend. The localization of 
$T_{max}$, which is the time scale when SF reaches plateau, is less 
clear in the cycle $\cal A$ than in the cycle $\cal B$, or in the
composite data.  During the cycle $\cal A$ there are no significant 
differences in the SF shape for all three UBV colors.  

The systematic difference between the cycles is also reflected in our
Monte Carlo simulations (see Table~\ref{tab:sffit}). We see most clearly 
a shift (reduction) in the maximum duration of the flares in the best fit 
model for the cycle $\cal A$ alone in comparison to the model for the whole 
Crimean data:  this reduction is from 800 days to 20 days.

%We did not make computations of the PSD for the two cycles 
%separately since the accuracy was too low to claim any systematic 
%evidences. 

%\begin{figure} 
%\vskip  3 truecm 
%\epsfxsize = 80 mm \epsfbox[50 180 560 660]{computations/Fig9.eps} 
%\epsfxsize = 90 mm 
%%\epsfbox[100 200 400 560]{sfb.eps} 
%\epsfbox{SF-CYCB.ps} 
%\caption{The structure function for the cycle B (Crimean data). 
%\label{fig:sfb}} 
%\end{figure} 
 
%****************************************************************begin{table*} 
\begin{table*}     
\caption{Structure function slopes b in various color bands, data sets,  
and timescales.
\label{tab:sfAB}}     
\begin{tabular}{|lcccccc}
\hline\hline     
Band~~~~~~~~~~~   &     $b$   &   $\sigma (b)$  &  $r$    &   $b$   &   $\sigma (b)$ &  $r$  \\     
\hline\hline      
\multicolumn{7}{l}{Cycle $\cal A$ (25.03.1968 - 03.07.1988)} \\    
  &  \multicolumn{3}{c}{(1 - 40) days}&\multicolumn{3}{c}{(80 - 3200) days}\\        
U      &    1.24 & 0.08 &  0.95  &  0.15 &  0.04 &  0.61 \\     
B      &    1.15 & 0.10 &  0.94  &  0.18 &  0.04 &  0.66 \\     
V      &    1.07 & 0.10 &  0.93  &  0.17 &  0.04 &  0.64 \\
mean UBV &  1.04 & 0.08 &  0.95  &  0.16 &  0.03 &  0.64 \\     
\hline     
\multicolumn{7}{l}{Cycle $\cal B$ (11.02.1989 - 13.03.2000)}\\    
&   \multicolumn{3}{c}{(1 - 40) days} & \multicolumn{3}{c}{(80 - 1500) days} \\       
U      &    1.24 & 0.05  & 0.98  &  0.58 &  0.04  & 0.94 \\     
B      &    1.12 & 0.03  & 0.99  &  0.64 &  0.04  & 0.96 \\     
V      &    0.86 & 0.04  & 0.98  &  0.67 &  0.04  & 0.96 \\     
mean UBV  & 1.02 & 0.04  & 0.99  &  0.63 &  0.04  & 0.96\\     
\hline     
\multicolumn{7}{l}{Cycles $\cal A+B$ (25.03.1968 - 13.03.2000)}\\    
&   \multicolumn{3}{c}{(1 - 50) days} & \multicolumn{3}{c}{(80 - 3200) days} \\        
U      &    1.21 & 0.08 &  0.96  &  0. 60 & 0.03 & 0.97 \\     
B      &    1.06 & 0.06 &  0.97  &  0. 64 & 0.03 & 0.98 \\     
V      &    0.95 & 0.05 &  0.97  &  0. 66 & 0.02 & 0.98 \\     
\hline     
\multicolumn{7}{l}{Photographic observations + Cycles $\cal A+B$ (5.03.1910 - 13.03.2000)}\\    
&   \multicolumn{3}{c}{(1 - 40) days} & \multicolumn{3}{c}{(80 - 3200) days} \\        
B      &    0.99 & 0.07 &  0.95  &  0. 60 & 0.03 & 0.98 \\        
\hline     
\multicolumn{7}{l}{X-rays (20.10.1974 - 05.01.2000)} \\
&   \multicolumn{3}{c}{80 min - 10 days} & \multicolumn{3}{c}{(10 - 4700) days} \\        
2-10 keV          &   1.24 & 0.04  &  0.98  &  0. 23 & 0.02 & 0.90 \\   
\hline 
\end{tabular} \\
Spectral bands are marked in column (1), time interval for 
correlated variability are specified  above columns (2 -- 4) and
above  columns (5 -- 7), slopes $b$ and errors of slope,
$\sigma(b)$, 
are shown in column (2), (3) and  (5), (6), and correlation 
coefficients $r$  for the corresponding linear regressions  are in column (4) 
and (7).  Errors of the slopes are determined from fitting the
power law to SF points in a specified timescale range.~~~~~~~~~~~~~~~~~~~~~~~~~~~~~~~~~~~~~~~~~~~~~~~~~~~~~~~~~~~~~~~~~~~~~~~~~~~~~~~~~~~~~~~~~~~~~~~~~~~~~~~~~~~~~~~~~~~~~~~~~~~~~~~~~~~~~~~~~~~~~~~~~~~~~~~~~~~~~~~~~~~~~~~~
\end{table*}     
%***************************************************************************** 
%88888888888888888888888888888888888888888
\begin{table}     
\caption{The SF model parameters determined from Monte Carlo simulations.
\label{tab:sffit}}     
\begin{tabular}{l|c|ccc}     
\hline\hline
Cycle & $\cal A$  & \multicolumn{3}{c}{$\cal {A + B}$} \\
\hline
Band & U   &   U  & B  & V  \\
\hline
dT [days]     & 0.10     &  0.20  & 0.25   &  0.25\\
$\beta$       & 0.45     &  0.25  &  0.25  &  0.17 \\
$\alpha$      & -1.47  & -1.80 & -1.70 & -1.60 \\
$T_{min}$ [days]     & 0.01   & 0.01 &   0.01 &   0.01 \\
$T_{max}$ [days]    & 20     & 800 &  800 &   800  \\
\hline
$\log(\tau)_{min}$ &   -0.05  &  0.03 & -0.05 &  0.03\\
$\log(\tau)_{max}$ &   1.3  &   3.0 &   3.0 &   3.0 \\
$b$   &   $1.22$ & $ 0.70$  &$ 0.79$&  $ 0.74$ \\
error $b$ (1$\sigma$)   &   $\pm0.17$ & $\pm0.07$  &$ \pm 0.06$&  $\pm 0.06$ \\
\hline     
\end{tabular} \\
The quoted values $dT, \alpha, \beta, T_{min}$ and $T_{max}$ are best fit model
parameters for flare separation, power law indices in the number of flares
and flare duration distributions, and the minimum and the maximum flare
duration. The slope $b$ of the SF was next determined in the time interval
between $\log(\tau)_{min}$ and $\log(\tau)_{max}$.
\end{table}     

%8888888888888888888888888

\subsubsection{X-ray band}\label{X-ray}

The shape of the SF in the X-ray band can be approximately represented
by a single power law with a slope $\sim 0.3$ (see Fig.~\ref{fig:SF}).
A flattening at the timescales above $\sim 1000 $ days is not excluded 
but it is not strongly required. Errors in this part of the plot are large.
Minor changes of the slope are also seen at $\log \tau$ around -1.8 and 0, 
and a complex structure is visible at timescales of 10 days.  

Because the scatter was so large, we reanalyzed the X-ray data as follows. 
We first created daily averages from all X-ray observations and
computed the long timescale part of the SF. Next we computed the short
timescale SF from the ASCA long monitoring alone, up to $\log \tau = 0.8$
(i.e. timescale corresponding to a half of the total duration of the
ASCA 2000 monitoring). We renormalized the SF dividing it by the variance,
(see Eq.~\ref{eq:NSF}). Both NSFs matched each other well at the transition
timescale. The results are shown in Fig.~\ref{fig:sim7}. The new plot 
appears smoother, but the overall shape did not change.

X-ray SF shows clearly the change of the slope around the timescale
of a few days and a trace of flattening at the longest timescales.
There is therefore a good correspondence between the SF and the NPSD
in the X-ray band. There is, however, a systematic difference between 
the SF in the X-ray band and in the optical band, hinted in our NPSD 
analysis. The difference is seen most clearly when all optical data 
are included (see the lowest panel of Fig.~\ref{fig:sim7} for a comparison 
with V band).  However, the difference is much smaller when only the 
photoelectric measurements from cycle $\cal A$ are taken (see the 
second panel of Fig.~\ref{fig:sim7}). In this case a significant 
difference is seen only at intermediate timescales from a fraction 
of a day to 10 days.  The presence of the two characteristic
timescales in X-ray band - $\sim 500$  days and $\sim 5 $ days -
should be confirmed by continued monitoring of this source.  

\subsection{Autocorrelation function}\label{acf}
 
\begin{figure} 
\vskip  3 truecm 
\epsfxsize = 90 mm 
%\epsfbox[100 200 400 560]{sfb.eps} 
%\epsfbox{ACF-U-X.PS}
\epsfbox{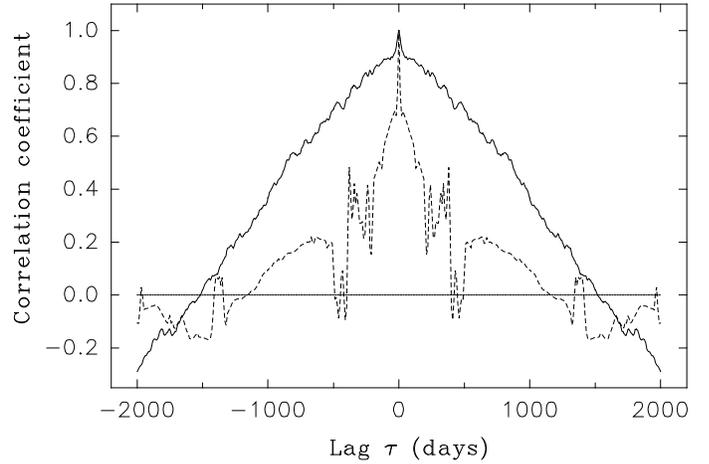} 
\caption{Autocorrelation function in U band (continuous line) and in the 
X-ray band (dashed line). 
\label{fig:auto}} 
\end{figure}

We supplement the analysis by calculating the autocorrelation function 
for the full data set in the U band and in the X-ray band. The 
difference between the two plots is remarkable. We see from 
Fig.~\ref{fig:auto} that X-ray flux shows very good correlation 
only at extremely short timescales below 10 days. At timescales of 
order of 10-1000 days, corresponding to a hump in a X-ray NPSD
(see  Fig.~\ref{fig:nupower}), a broad shoulder replaces the sharp peak and 
falls down slowly. A significant minimum in the correlation is seen at
a timescale of $\sim 215$ days, which, when multiplied by 2, corresponds 
well to the first minimum in NPSD. It is consistent with the SF not 
being completely flat at such a long timescales (see Table~\ref{tab:sfAB}). 
The correlation reaches zero at 400 days, giving timescales of 800 days, 
in agreement with the presence of considerable power in NPSD down 
to timescales of $\sim 1000$ days. 

U band autocorrelation function does not show a 
similar structure and falls down steadily, crossing 0 at a time lag 
of 1650 days and indicating the characteristic timescale of about 3300 days.
This is again in rough agreement with the determination from SF
($\sim 3200$ days) and NPSD ($\sim 5000$ days). 
 
\section{Discussion}\label{discussion}
  
\subsection{Character of optical and X-ray variability}\label{acc}

Our results indicate that there are two variability mechanisms operating in
the Seyfert galaxy NGC~4151, and we infer this by comparing the results of 
our analysis of the optical and X-ray data.  

The NPSD plots for the optical band suggest the 
flattening at the timescales of $\sim 10$ years and Monte Carlo
simulations show that this value is actually a lower limit for the
characteristic timescale. The SF plots for the optical band suggest a 
flattening at $\sim 4 - 10$ years and the Monte Carlo simulations show that
models with shot duration shorter than 800 days cannot explain the data.

On the other hand, the NPSD plots for the X-ray band suggest that most of the
power is present in the intermediate timescale range, between 5 and 1000 days.
Monte Carlo simulations support this view, giving the best fit flattening
timescale at 130 days and a 1 $\sigma$ error upper limit of 5 years. SF plots
in the X-ray band show that the power rises mostly at 1 - 10 day timescale
range. The distribution is rather flat at longer timescales, with a hint of 
flattening above 1000 - 2000 days.  

In summary, although errors are large, we conclude that the optical variability
occurs predominantly on timescales longer than $\sim 10$ years while the
X-ray variability is mostly confined to the timescale range 5 $ {\rm days} -
3$ years. Therefore, two different variability mechanisms are apparently at
work.  Our results therefore support the early claim of Belokon, 
Babadzhanjantz, \& Lyutyi (1978) based on much more limited data.

The difference in the dominating timescales in the optical and X-ray band 
does not exclude some interaction between the two bands.  Long time scale 
behavior of the variability of X-ray flux appears to be somewhat 
correlated with the optical flux, suggesting an influence of the 
structure of the optical emitter on the X-ray emitter. Also, 
the fast variability, similar to the X-ray variability, is present in the
optical data which possibly reflects the effect of the X-ray irradiation 
of the cool material.  

The shortest characteristic timescale seen in our X-ray data is in
agreement with the timescale $13^{+11}_{-5}$ days determined by 
Collier and Peterson (2001). We do not see clearly any change of 
the NPSD slope if we try to extend our computations towards shorter
timescales but the SF shows a change of the slope just above this range.
Collier and Peterson (2001) did not find the long timescale component 
because their data for NGC~4151 covered only 99 days. Their conclusion
about the overall similarity between the optical/UV variability and X-ray
variability {\it on short timescales} agrees with our results, although 
our results are not particularly accurate in that range.

Our inference that the long timescales play  the dominant role of 
the optical band cannot be an artifact of the complex analysis since 
it is actually apparent from Fig.~\ref{fig:Bcurve}.  The actual value 
of the characteristic timescale is strongly influenced by the
recent outburst (cycle $\cal B$) lasting from 1989 till 2000, although 
hints for similar timescales are present in the older data as well, 
when they are superimposed on low amplitude longer trend.

The lack of trends longer than $\sim 3 $ years in the X-ray data 
obtained from NPSD is less secure and a clear absence of longer timescales 
is also not immediately apparent from the SF analysis.  Large errors 
associated with the early measurements make this determination rather 
difficult. A few more years of further monitoring will resolve this
issue definitively. However, the dominant role of much shorter timescales
in the X-ray band seems to be well established.  

\subsection{Correlation of optical and X-ray luminosity}\label{corr}

Long data sets provided also an opportunity to better study the correlation
between the U band and the 2 - 10 keV X-ray emission. The correlation 
between the whole data sets is surprisingly poor.
Short timescale study of UV and X-ray variability of Edelson et al.(1996)
showed a very strong coupling during 10 days of monitoring. Noticeable 
correlation was found by Perotti et al. (1990) between the hard X-ray 
35 keV flux and the optical flux.  

Oknyanskij (1994) discussed the issue of the dependence of broad band 
correlations on the luminosity state of the nucleus and he found  that 
in the low state in 1983 - 1986, correlations are more significant than 
in the high state in 1975 - 1978. The same conclusion was drawn by Perola
\& Piro (1994).

\begin{figure} 
\epsfxsize = 90 mm 
\epsfbox{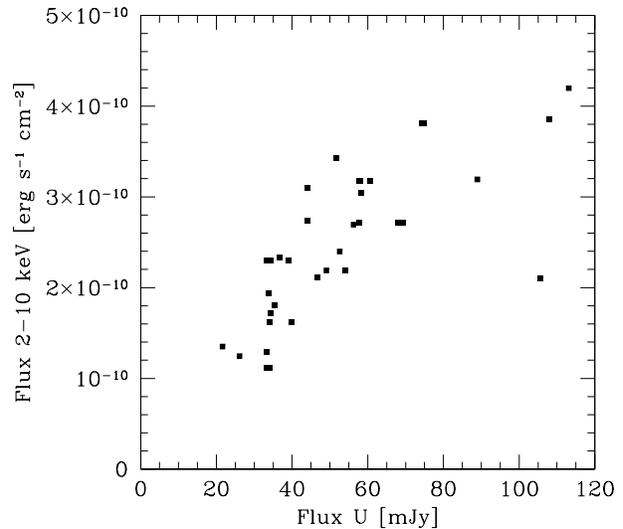} 
\caption{The correlation between the optical flux in U band and 
the X-ray flux for the data points collected starting from 1990, with U and 
X measurement separation smaller than 1 day. 
\label{fig:corUX}} 
\end{figure}

X-ray and U band measurements in our data sets are frequently significantly 
separated in time and the old X-ray data are not very accurate. Therefore, we 
repeated our analysis for a subset of data points including only the relatively
new and accurate measurements (observations starting from 1990) and taking
only those pairs for which the separation of the measurement time is smaller
than 1 day. Our data sets contain 37 such pairs. The result is shown
in  Fig.~\ref{fig:corUX}. The distribution observed in the plot is consistent 
with the coefficient of correlation
$r=0.780$, and is highly inconsistent with the null hypothesis $H_0$
assuming no correlation, i.e. $r=0$. We tested it following the procedure of
Borczyk,Schwarzenberg-Czerny \& Szkody (2003). For $n=35$ degrees of freedom,
our value of $r$ tested against $H_0$ yields Student $t=7.36$
corresponding to the significance level $1-2.6\cdot10^{-8}$. 
%(two-sided test, corresponding to $5.5 \sigma$).

The result strongly confirms the trend seen by Oknyanskij 
(1994) and Perola \& Piro (1994): the correlation is stronger when the source
is fainter, and the X-ray flux shows significant systematic 'deficit' when 
the source is bright. It shows that the smaller X-ray rms variability (see
Table~\ref{tab:rms}) does not result from the presence of a constant 
X-ray component. In opposite, we rather have a small constant U component 
(offset), a linear trend for moderate flux and a flattening for high flux,
but with significant dispersion. 

\subsection{Comparison with NGC 5548}\label{compar}

Although the power spectra in the X-ray band are available 
now for a number of AGN, the variability in the
optical band was quantitatively studied only for
the Seyfert galaxy NGC~5548. Therefore, we can only compare 
the results for the two Seyfert galaxies: NGC~4151 and NGC~5548. 

Below, we compare the structure functions in U, B, V, $\mathrm{R_J}$ 
bands of NGC~4151 with the results for NGC~5548 obtained by 
Doroshenko et al. (2001). Our result is that the SF of NGC~4151 
in U and B bands has a break (see Table~\ref{tab:sfAB} and 
Figures~\ref{fig:SF}, \ref{fig:sfa}), 
%\ref{fig:sfb})
whereas SF of NGC~5548 has a single slope in all bands.  
The values of the slope, $b$, are 0.72 $\pm$ 0.02 for U band and 
0.69 $\pm$ 0.02 for B band roughly similar to our results for the 
long timescale behavior in Cycles $\cal A+B$. However, the timescale where 
the SF flattens may be somewhat different in the two objects: in NGC~4151 
it is about $\log \tau \approx 3.2 - 3.6$ [day] while in NGC~5548 it is 
somewhat shorter, just below $\log \tau = 3.0$ [day]. Note, however, 
that the slopes for NGC~5548 were determined without correcting for 
the level of the observational error.
%{\bf jak mozna tu wprowadzic 
%poprawki na bledy obserwacyjne?  czy to znaczy ze 
%bledy nie zostaly poprawnie wyznaczone?  Nie bardzo rozumiem}

We can also compare the normalized power spectrum density of NGC~4151 
in the optical band (Fig.~\ref{fig:nupower}) with NGC~5548 given by 
Czerny et al. (1999) in their Fig.~7. We can see an apparent maximum in 
$Power \times Frequency$ representation in both objects, at 
$\log f \simeq  -3.7$ [1/day] for NGC~4151 and at $\log f \simeq -3$ 
for NGC~5548. However, Monte Carlo simulations show that for NGC~4151 
we actually see a lower limit of the dominant timescale and the same 
can be true for NGC~5548 with data set covering shorter period.

We also compare the normalized power spectrum density  of NGC~4151 in the 
X-ray (2 - 10 keV) band (Fig.~\ref{fig:nupower}) with NGC~5548 given by 
Uttley et al. (2002).
The overall shape of PSD in both objects is similar.  
The PSD of NGC~5548 is well represented by a single power law 
($\alpha = 1.6$) up to log $f$ = -3. For NGC~4151 the fit is better for a 
solution with a break ($\alpha = 1.5$, log $f$ = -2.1) but a single 
power law is also acceptable.

The masses of the central black holes estimated from reverberation 
studies are $\log M=7.83$ for NGC~5548 (Peterson \& Wandel 1999)
and $\log M = 7.08$ for NGC~4151 (Wandel, Peterson, \& Malkan 1999).  
Mass determination from accretion disk fitting gives log $M$ = 7.4 
for NGC~5548 and log $M$ = 7.0 - 7.3 for NGC~4151 (Czerny et al. 2001). 
The high frequency tail of our NGC~4151 power spectrum 
is shifted down by a factor of 1.5 in comparison with the NGC~5548 
power spectrum of Uttley et al. (2002).  Scaling used by Hayashida 
et al. (1998) in such a case implies that the black hole mass in 
NGC~4151 is higher, in contrast to the above results.   This suggests 
that indeed the black hole masses in NGC~4151 and in NGC~5548 are 
comparable, but again, the errors are large, probably higher than 
quoted in the above papers.

\subsection{Comparison with Cyg X-1}\label{cyg}

\begin{figure} 
\epsfxsize = 90 mm 
\epsfbox{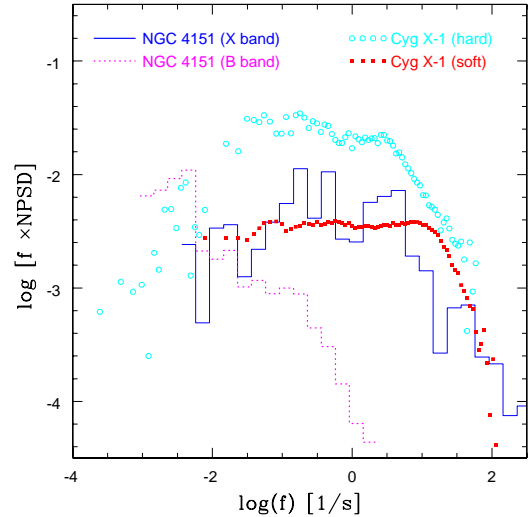} 
\caption{The NPSD for Cyg X-1 in the soft state, June 1996, 
(filled squares, from Gilfanov et al. 2000) and in the hard 
state February 3, 1997, (open circles), together with the NPSD 
for NGC~4151 in the optical band (dotted histogram) and in the 
X-ray band (continuous histogram), both shifted horizontally by 6.3
to account for a mass difference between the two sources. 
\label{fig:cygx1}} 
\end{figure} 

The comparison of observational characteristics of 
galactic black hole candidates (GBH) and AGN is not straightforward.  
Observed photon spectra generally consist of two 
principal spectral components: soft bump usually interpreted as the
accretion disk emission, and a power law interpreted as a result of
Comptonization by a hot optically thin plasma. In AGN these components 
are well separated: the first one dominates the optical/UV emission, 
while the second one dominates the X-ray band.  In galactic sources 
both are seen in X-ray band but their dominant role depends on the 
luminosity state of the source.  Therefore, the power spectrum 
of a GBH in the hard, power-law dominated state should be compared 
with the X-ray power spectrum of an AGN while the power spectrum of
a GBH in the soft disk-dominated state should be compared with
the optical power spectrum of an AGN.  While the former comparison 
has been attempted often in the past, the latter has not been addressed 
extensively.  

X-ray power spectra of AGN are generally quite similar to power spectra of
galactic objects in their hard state (see e.g. Czerny et al. 2001 and the
references therein). The power peak in $f \times NPSD$ diagram is broad,
covering about 2 decades, and its position on the frequency axis depends
linearly on the mass of the black hole and can 
well be used for the mass determination
(Hayashida et al. 1998; Czerny et al. 2001). A comparison of the 
X-ray power spectra of Cyg X-1 and NGC~4151 (Fig.~\ref{fig:cygx1}) 
generally support this picture, and suggest that the mechanisms 
responsible for the emission and variability of the power law spectral 
component are indeed identical for AGN and GBH.  

However, a comparison of the optical power spectrum of an AGN 
with power spectra of galactic sources in soft state does not
show any striking similarity. In galactic sources the power is
evenly distributed across the broad range of frequencies. In AGN, 
the longest timescales dominate. Our results for NGC~4151 confirm 
such a view (see Fig.~\ref{fig:cygx1}). Timescales dominating 
the optical/UV variability in AGN (years) correspond to the timescales 
of hundreds/thousands of seconds in galactic sources. Strong outbursts with  
such timescales are only seen in the microquasar GRS1915+105 (see Belloni et 
al. 1997, Janiuk, Czerny, \& Siemiginowska 2000).  However, this 
source shows strong jet outflow and therefore it is not a simple 
analogue of a radio quiet Seyfert galaxy. 

\subsection{Decomposition of optical light curve into fast and slow variability
components}\label{two}
 
The presence of the two separate types of variability - a slow 
component and a fast component - was already suggested by Belokon et al. 
(1978). Our results from the analysis of the PSD and SF in U and B bands 
based on much better data coverage confirm this result.  Therefore, 
the data are not well represented by a sum of a constant component 
and a rapidly variable component, as suggested by Ulrich (2000). 

\begin{figure*} 
\epsfxsize = 160 mm 
%\epsfbox{/home/bcz/4151opt/computations/comparison2.eps}
\epsfbox{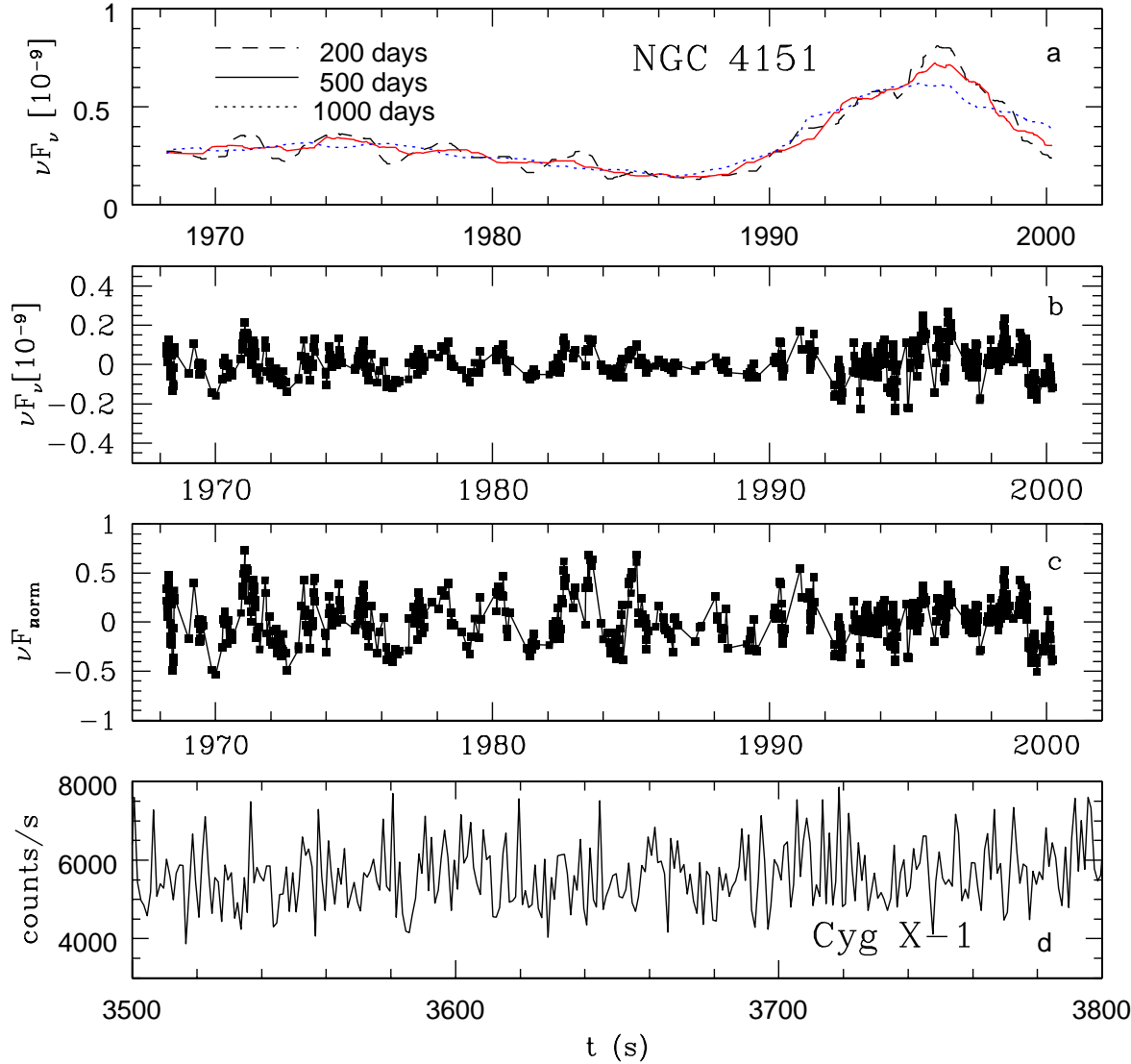} 
\caption{Panel (a): 
the smoothed light curve of NGC~4151 in U band, where smoothing 
timescales equal 200, 500 and 1000 days were used.  
Panel (b): the light curve of 
NGC~4151 in U band, with the long timescale  trend of 500 days 
subtracted. Panel (c): the normalized light curve of NGC~4151 in U 
band, with the long timescale trend of 500 days subtracted.  Panel 
(d): X-ray light curve of Cyg X-1 when the source was in the 
soft state (May 30, 1996). Errors in the Crimean data are $\sim 2.2$ \%,
errors in Cyg X-1 light curve are $\sim $ 25 cts s$^{-1}$. 
\label{fig:comparison}} 
\end{figure*}

In order to visualize the presence of the two separate types of
variability -- a slow component and a fast component-- we complement 
the computations of the optical NPSD and the SF with a simple illustration. 
We calculate the smooth long timescale light curve at each data 
point by averaging all available data points separated from a 
given instance by less than adopted smoothing timescale, $\Delta T$. 
A representative set of such light curves for a range of $\Delta T$ is 
shown in Fig.~\ref{fig:comparison}. 

%\begin{figure} 
%\epsfxsize = 80 mm \epsfbox[50 180 560 660]{computations/Fig9.eps} 
%\epsfxsize = 90 mm 
%\epsfbox{fig1_tot_seq.eps} 
%\caption{The smoothed light curve of NGC~4151 in U band, for the smoothing 
%timescale equal 200, 500 and 1000 days. Upper and lower light curve are 
%shifted 
%by 0.2 for clear display. 
%\label{fig:long}} 
%\end{figure} 

Adopting  $\Delta T = 500$, we display the short timescale variations 
by subtracting the smoothed light curve from the original one (see 
Fig.~\ref{fig:comparison}, panel (a)). The result is shown
in Fig.~\ref{fig:comparison}, panel (b). We see that the amplitude of 
variations display significant trends.  This dependence of the amplitude 
on the flux was discussed by Doroshenko et al. (2001). 

Such trends almost disappear if we plot the normalized light curve, 
i.e. if the value of the flux difference is divided by the value 
of the flux of the smoothed light curve (see Fig.~\ref{fig:comparison}, 
panel (c)).  This figure provides an illustration of the same fact 
that was stressed by Uttley \& McHardy (2001): only the normalized power 
spectrum is independent from the luminosity state of the source.

We can now compare these trends with variability of Cyg X-1 in its 
soft (disk-dominated) spectral state. In Fig.~\ref{fig:comparison} 
(panel (d)) we show a sequence from the light curve of Cyg X-1 in 
February 3, 1997 (30 yr for NGC~4151 are equivalent to 300 s of 
Cyg X-1). We see that this light curve is clearly different 
from the original U-band light curve of NGC~4151 (see Fig.~\ref{fig:UXcurve}) 
but there is some similarity of Cyg X-1 light curve to that 
shown in panel (c) where all the long trends from NGC~4151 
light curve have been removed.  This supports our earlier conclusion 
that the fast optical variability component in AGN may well be 
connected with the X-ray reprocessing, and the X-ray variability 
itself is similar in AGN and GBH so the nature of fast variability 
might be the same in both types of objects.
 
%\begin{figure} 
%\epsfxsize = 80 mm \epsfbox[50 180 560 660]{computations/Fig9.eps} 
%\epsfxsize = 90 mm 
%\epsfbox{fig1_diff_nufnu_norm.eps} 
%\caption{The normalized 
%light curve of NGC~4151 in U band, with the long timescale 
%trend of 500 days subtracted. 
%\label{fig:short_norm}} 
%\end{figure} 

\subsection{Intrinsic variability versus obscuration}\label{obscur}

A number of authors suggested that the variable obscuration is, at
least partially, responsible for the observed variability of AGN
(e.g. Collin et al. 1996, Boller et al. 1997, Brandt et al. 1999, 
Abrassart \& Czerny 2000, Risaliti, Elvis, \& Nicastro 2002). 

The absorbing column density in Ginga/EXOSAT data seems to be 
lower when the source is brighter (Yaqoob \& Warwick 1991). However, 
if we look at the most recent and accurate observations, no obvious 
correlation is seen. In December 1993 when the AGN Watch monitoring 
was done the source was bright in U band ($\sim 75$ mJy; Kaspi  et al. 
1996, Crimean data) and the intrinsic X-ray column was $3.5 \times 
10^{22}$ cm$^{-2}$ (Edelson et al. 1996). In the beginning of 
March 2000 the source was faint in the U band ($\sim 21.5$ mJy; 
Crimean data) and the intrinsic X-ray column determined from
the Chandra data was $3.7 \times 10^{22}$ cm$^{-2}$ (Ogle et al. 2000). 

On the other hand, the observed delays between the optical continuum and 
the emission lines (e.g. Kaspi et al. 1996) and between the optical 
emission and the infrared emission (e.g. Oknyanskij et al. 1999) show 
that the variability on the timescales of days is certainly intrinsic 
to the source, at least partially.  

IUE monitoring during 18 years, which covered both the very low state 
around 1988 and the luminosity peak in the middle of outburst $\cal B$ 
around 1995 showed a factor over 12 increase in the flux although the 
UV line CIV1550 showed lower amplitude, by a factor of 5 (see Fig. 11 
of Ulrich 2000).  It might suggest that the continuum variations
are additionally enhanced by some variations in the absorbing 
medium, perhaps due to variable ionization. 

This shows that although the variable obscuration cannot be totally excluded, 
it cannot be the only factor responsible for the observed variability 
and most of variations are instead intrinsic to the source.
 
\subsection{Interpretation of the variability}\label{instab}
 
The strong variability of AGN in optical, UV and X-ray band is a well known 
but not well understood phenomenon (see Mushotzky, Done, \& Pounds 1993;  
Ulrich, Maraschi, \& Urry 1997).  The observed properties of the X-ray 
time series indicate some stochastic process (McHardy \& Czerny 1987;  
Czerny \& Lehto 1997), which physically can be interpreted as magnetic 
flares above the cold disk or shocks developing in the hot accreting
material. Optical variations are partially caused by reprocessing (although
modeling is not simple; see e.g. Rokaki, Collin-Souffrin, \& Magnan 1993;  
Kazanas \& Nayakshin 2001;  \. Zycki \& R\' o\. za\' nska 2001;  
Wang, Wang, \& Zhou 2001) and partially by another process operating 
on longer timescale, possibly the radiation pressure instability 
(e.g. Czerny et al. 1999). 

The variability of the flux from the cold disk can be related to the 
Keplerian timescale, $t_K$, thermal timescale, $t_{th}$  or viscous 
timescale $t_{visc}$ of the accretion flow. The first value 
is universal (and depends only on the mass of the central object), 
and the other two can be computed for a Shakura-Sunyaev disk model.  
The variations due to the changes of properties of the hot plasma may 
also be related to $t_K$ and $t_{th}$, although the lack of theory for 
the dynamics of the hot phase makes this parameterization less firm.  

To compute these time scales, we fix the mass of the black hole at 
$3.7 \times 10^7 M_{\odot}$, based on variability studies,  
and take the average luminosity to the Eddington luminosity 
ratio 0.02 resulting from the black hole mass value and the estimate of the
total bolometric luminosity (see Czerny et al. 2001). For those 
parameters, we obtain the radial dependences of those timescales:
\begin{equation}
t_K= 0.18 m_{3.7} r_{10}^{3/2} ~~~ {\rm [d]},
\end{equation}
\begin{equation}
t_{th} = 1.8 \alpha_{0.1}^{-1}m_{3.7}r_{10}^{3/2} ~~~{\rm [d]},
\end{equation}
\begin{equation}
\label{eq:visc}
t_{visc} = 4000 \alpha_{0.1}^{-1}m_{3.7}r_{10}^{7/2}\dot m_{0.02}^{-2}  ~~~{\rm [d]},
\end{equation}
where $r$ is expressed in units of 10 Schwarzschild radii, $m$
in units of  $3.7 \times 10^7 M_{\odot}$, and a standard viscosity 
parameter $\alpha$ in units of 0.1. The mass of the black hole enters 
the formulae linearly, and the accretion rate affects only the third 
quantity. In these computations we adopted $\dot M_{Edd}= 1.69 \times 
10^{18} \times (M/M_{\odot})$ g s$^{-1}$, and $L_{Edd}= 1.27 \times 
10^{38} \times (M/M_{\odot})$ erg s$^{-1}$, corresponding to the Newtonian 
accretion efficiency of 1/12. 

We can now compare those timescales with the characteristic points 
present in the NPSD and SF both in the optical and in the X-ray band.
The derived timescales are related to the Fourier frequencies: 
$f = (2 \pi t)^{-1}$. The factor $2 \pi$ is important if we want to 
compare the predicted timescales with the characteristic
frequencies in the shape of the NPSD. Therefore, observed characteristic
frequencies (1/5, 1/1000 and a lower limit of 1/4000 d$^{-1}$) translate
into 0.8 d, 160 d, and more than 640 d, respectively. Here we used 
the timescales resulting from two-break fit (see Table~\ref{tab:psfit}), 
motivated by the analogy with the galactic sources. 
 
%\begin{figure} 
%\epsfxsize = 80 mm \epsfbox[50 180 560 660]{computations/Fig9.eps} 
%\epsfxsize = 80 mm 
%\epsfbox{mdot_rout.eps} 
%\caption{The dependence of the extension of the thermally and viscously
%unstable region on the accretion rate $\dot m$ for black hole mass
%$2 \times 10^7 M_{\odot}$ and $\alpha=0.1$.
%\label{fig:rout}} 
%\end{figure} 

%\begin{figure} 
%\epsfxsize = 80 mm \epsfbox[50 180 560 660]{computations/Fig9.eps} 
%\epsfxsize = 80 mm 
%\epsfbox{ms10_15.eps} 
%\caption{The stability curve (relation between the disk surface density
%$\Sigma$ and the accretion rate $\dot m$) for two disk radii,
% black hole mass $2 \times 10^7 M_{\odot}$ and $\alpha=0.1$. The complex
%shape is responsible for non-monotonic relation seen in Fig.~\ref{fig:rout}.
%\label{fig:stabcurve}} 
%\end{figure}

The shortest X-ray timescale corresponds to the thermal timescale at the 
marginally stable orbit of the non-rotating black hole 
if the viscosity parameter is $\sim 0.03$, as obtained 
from spectral fits to the data for NGC~5548 (Kuraszkiewicz, Loska, \& 
Czerny 1997). The value of the longest X-ray timescale is by a factor of 
200 longer which suggests that the X-ray production region extends in 
that case up to $\sim 100 R_{Schw}$, if timescale scaling as $r^{3/2}$ holds. 

The timescale dominating the optical variability seems to be too long 
for pure thermal variations related to the radiation pressure 
instability.  The thermal timescale of 640 d corresponds to a distance 
of $\sim 220$ $R_{Schw}$ (assuming $\alpha$ as above), much larger than 
the extension of the instability zone ($\sim 70$ $R_{Schw}$, calculated 
using 
the code of R\' o\. a\' nska et al. 1999) for the adopted system 
parameters. Viscous timescales (see Eq.~\ref{eq:visc}) on the other 
hand, are much longer than the period covered by the optical data.  
However, recent computations performed by Szuszkiewicz (1999) indicated 
much faster evolution than estimated from Eq.~\ref{eq:visc} - 
an outburst lasted $\sim 11 $ years for adopted model parameters: $10^8 M_{\odot}$
black hole, $\alpha = 0.1$ and accretion rate $\dot m = 0.06$.

Therefore, the observed variability in the optical band neither 
obviously proves nor contradicts the supposition that the optical 
variability at long timescales is caused by the radiation pressure 
instability of the optically thick disk.  Further optical monitoring 
as well as theoretical progress are needed to resolve this issue. 
However, the lack of strong long timescale (hundreds of seconds) 
variability and the lack of radiation pressure dominated zone in  
Cyg X-1 (see e.g. Gierli\' nski et al. 1999;  Janiuk, Czerny, \& 
Siemiginowska 2002), when compared with the presence of long timescale 
variations and the disk unstable zone in NGC~4151, 
favors the radiation pressure scenario. Analysis of the optical 
variability of NGC~5548 supported this view as well (Czerny et al. 1999).  

\section{Conclusions}

Our analysis of the 90 years of the optical data and 27 years of the 
X-ray data for NGC~4151 using the NPSD and the SF techniques, gives 
the following results:

\begin{itemize}

\item {variability properties in the optical and X-ray band are 
significantly different}

\item {X-ray variations are predominantly in the frequency range
1/5 - 1/1000 $d^{-1}$}

\item {optical variations are dominated by a component with a
frequency on the order of (or possible smaller than) $\sim 1/5000$  
d$^{-1}$ (or $\sim 10$ years), but the short timescale variability 
may be related to X-ray variability}

\item {the presence of long timescale variability in NGC~4151 and 
the absence of analogous variability (on timescales of hundreds of 
seconds) in Cyg X-1 favors the radiation pressure mechanism}

\item {the radial extension of the X-ray production region is similar 
in NGC~4151 and Cyg X-1 so the possible presence of radiation pressure 
dominated region in NGC~4151 and its absence in Cyg X-1 does not 
affect the X-ray production}

\end{itemize}

\section*{Acknowledgements} 

We are grateful to an anonymous referee for many helpful remarks which
led to significant improvement of the manuscript. 
We thank S.G. Sergeev for his permission to use the software 
package for data analysis.
% and J. Lehar for making the code {\sc clean}  publicly available.
We also acknowledge the help with Cyg X-1 data and 
many enlightening discussions with Piotr \. Zycki. This work was 
supported in part by grants 2P03D~003~22 (BCz \& ZL), and 
5P03D~002~20 (ASC) %(ASC *** to jest juz moj poprawny numer) 
of the Polish State 
Committee for Scientific Research (KBN), as well as NASA Chandra grants 
awarded by SAO to SLAC as GO0-1038A and GO1-2113X.  V.T. Doroshenko has 
also benefited from the support from the Russian
Basic Research Foundation Grant N00-02-16772a and Federal
Target Scientific-Technical Program, section "Astronomy"
2002-2006.

\section*{Appendix A: Determination of the error on the power spectrum}

Two methods can be used to estimate the  errors on the power spectrum
determined from observations. The first method is to assign directly the
errors on the plotted histogram, and the second one is to do so through the 
Monte Carlo simulations. In this paper, we use both methods.

\subsection*{A.1 The direct method}

The direct measurement error is estimated as follows.  
Discrete power spectrum computed at Fast Fourier Transform (FFT) 
frequency grid has a $\chi^2(2)$ distribution. Its variance matches 
the expected power. Whether the same is true for unevenly sampled 
data remains yet to be proven (e.g. Timmer \& K\" onig 1995).  Below, 
we adopt the same procedure, formal objections notwithstanding. Since 
we use logarithmic values (see Papadakis \& Lawrence 1993), we estimate
the error on the logarithmic value as follows. If the true value
of the power is distributed around the measured value uniformly between 0
and twice the value, one sigma negative error would correspond to a value
equal to 0.16 of the measured value. Therefore, the error of the logarithmic
value on the negative side is equal to 0.43. We assume now that the errors are
symmetric and adopt this value as a single measurement error of a single power
spectrum in a single bin.  

The power spectrum is computed in the frequency range determined by the data.
It is subsequently binned into logarithmic bins of the width of 0.2.
Measurements grupped into a single bin usually show significant dispersion,
sometimes even higher than the single measurement error assigned above.
Therefore we compute the error of a power spectrum in a bin by combining
single measurement errors with the dispersion around the mean value
$\bar{x}$
\begin{equation}
\sigma _b (f_k) = ({\sum_i {[x_i(f_k) - \overline{x(f_k)}]^2 \over n_k^2}} +
{0.43^2 \over n_k^2})^{1/2}.
\end{equation}

When separate sequences are computed overlapping in the frequency range we
add the spectra in the common frequency bins assuming that the weight of the
spectrum is inversely proportional to the measurement error of the contributing
spectrum. The error of the resulting spectrum is again computed by combining
the errors of the single measurements $\sigma _b^i$ (determined as described
above) with the dispersion error

\begin{equation}
\sigma _b (f_k) = ({\sum_i {[x_i(f_k) - \overline{x(f_k)}]^2 \over n_k^2}} +
{\sum_i {(\sigma _b^i)^2 \over n_k^2})^{1/2}}.
\end{equation}

This method is very simple and does not depend on the model of the power 
spectrum. However, it does not incorporate well the systematic errors
caused by the window function.  

\subsection*{A.2 Monte Carlo simulations of light curves}

This method allows to estimate also the systematic effects of the 
window function as well as to draw the error contours for the adopted 
parameterization of the shape of the power spectrum.

In our approach we follow the method used by Uttley et al. (2002) and 
based on the method of Timmer \& K\" onig (1995).  It is equivalent to the
method used by us in our previous paper (Czerny et al. 1999) which consisted
of filtering the white noise with the filter corresponding to the 
adopted power spectrum distribution.

In our specific case we proceed as follows:  

\noindent $\bullet$ We assume a shape of the NPSD in a form of either 
a broken power law with the slope 0 below the frequency break 
(parameters: frequency break and the slope above it) or in the form 
of a power law with two breaks and the slopes 0, 1 and 2 
(parameters: two frequency breaks).

\noindent $\bullet$ We generate a random realization of this distribution 
by assuming the random uniform distribution of the power spectrum 
around the adopted value and the random distribution of Fourier phases.
We generate an artificial equally spaced light curve that is somewhat 
longer than the duration of the data set through the FFT method. 
Next the light curve is folded with the observational window function 
thus reducing the long light curve to the number of points and spacing 
as in the observational data set.

\noindent $\bullet$ Next, we follow exactly the procedure applied to the 
observational data, i.e. we form yearly averages and other data subsets 
(see Sect.~\ref{sect:metpower}). We compute the power spectrum 
$NPSD_i(f)$ as we did for the observational data.  We repeat the
entire procedure $i = 100 - 500$ times in order to create good statistics.
The method is computationally intensive, so we cannot significantly 
increase this number for such a long and densely sampled data sets 
as used in our paper. From this distribution, we determine the average 
simulated power spectrum $\overline{NPSD(f)}$, and the dispersion 
$\sigma(\overline{NPSD(f)})$. We can assign a $\chi^2_{dist}(i)$
value to each of the simulated light curves:  with the partial
results forming the $\chi^2$ statistics around it 
\begin{equation}
\chi^2_{dist}(i)=
\sum_{k=1}^{kmax} {[NPSD^i(f_k) - \overline{NPSD(f_k)}]^2
\over \sigma(\overline{NPSD(f_k)})^2}.
\label{eq:MCchipow}
\end{equation}
The normalization of the power spectrum is adjusted to minimize the
$\chi^2_{dist}(i)$ value.

\noindent $\bullet$ We also compute the same $\chi^2_{dist}(obs)$ value 
for the mean simulated power spectrum and the power spectrum determined 
from observations:  
\begin{equation}
\chi^2_{dist}(obs)=
\sum_{k=1}^{kmax} {[NPSD^{obs}(f_k) - \overline{NPSD(f_k)}]^2\over \sigma(\overline{NPSD(f_k)})^2},
\end{equation}

\noindent $\bullet$ The quality of the fit for a given power spectrum 
parameters is determined from the comparison of the value 
$\chi^2_{dist}(obs)$ with the distribution given by 
Eq.~\ref{eq:MCchipow}: the percentage of the light curves
with the $\chi ^2_{dist}(i)$ smaller than $\chi^2_{dist}(obs)$ gives the
probability $P$ of the model rejection.

\noindent $\bullet$ Repeating the analysis for the range of the power 
spectrum parameters we find the best fit value (the highest acceptance 
probability) and we check if it is an acceptable model. 

\noindent $\bullet$ Next, we determine the error contours for the two 
parameters of interest. For that purpose, we assume that the probability of 
any sets of values is proportional to the acceptance probability 
obtained above. We normalize the probability through performing the 
two-dimensional integral. We obtain the contour errors corresponding 
to the rejection probability of 68\% ($\sim 1 \sigma$ error) and 
90\% by integrating this renormalized probability over the surface 
through subsequent probability contours until the value of the integral 
is equal to 0.68 or 0.9, correspondingly. Since 
the computations were performed on a regular grid, in practice we
find for example the contour level $P_{0.68}$ on the parameters of 
interest $x$ and $y$ iteratively from the expression
\begin{equation}
0.68 = {\sum_{i,k}^{if ~P >P_{0.68}} P(x_i,y_k) \over \sum_{i,k} P(x_i,y_k)}.
\end{equation}
Such contours are slightly smaller than contours plotted directly from the
acceptance probability. 

\begin{figure} 
\epsfxsize = 90 mm 
\epsfbox{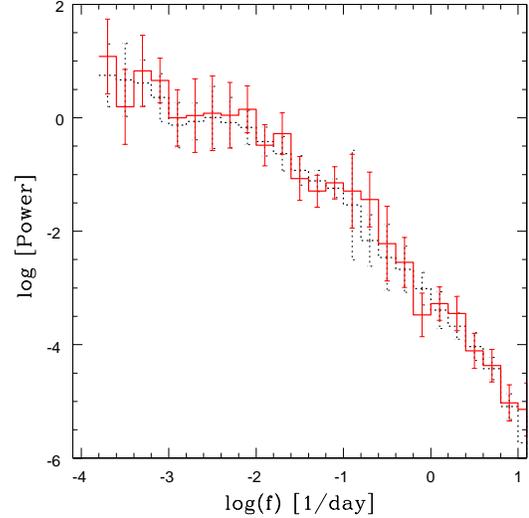} 
\caption{Observed NPSD in the X-ray band (continuous histogram) with errors
determined directly from the data plotted against the best fit model (dotted histogram) of a
single break
power law with $\log f_0 = -2.1$, $\alpha = 1.5$. The  $\chi ^2_{dist}(obs)$ for this model is 13.39 for 22 dof (25 frequency points, 3 model parameters including normalization).
\label{fig:bestfit}} 
\end{figure} 

The comparison of the observed NPSD with the best fit model in X-ray band is 
shown in Fig.~\ref{fig:bestfit}. The two distributions agree well within the 
indicated errors.

\subsubsection{Comparison with standard $chi^2$ statistics}

\begin{figure} 
\epsfxsize = 90 mm 
\epsfbox{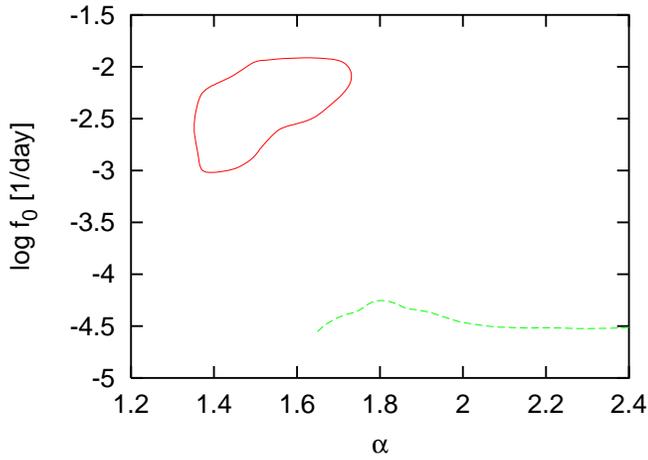} 
\caption{Error contour at 90\% confidence level for the broken
power law model of X-ray and optical NPSD determined as 
$\chi^2_{min} + 4.61$; to be compared with thin lines in
Fig.~\ref{fig:err1}, upper panel.
\label{fig:chi2_46}} 
\end{figure}

\begin{figure} 
\epsfxsize = 90 mm 
\epsfbox{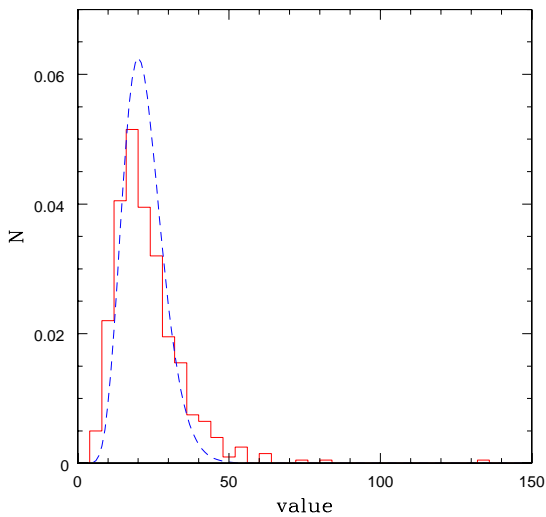} 
\caption{The distribution of the $\chi ^2_{dist}(i)$ values (histogram) for
500 Monte Carlo curves in comparison with the $\chi ^2$ theoretical
plot for 22 dof. Parameters: X-ray data, model of a single break
power law with the best fit parameters: $\log f_0 = -2.1$, $\alpha = 1.5$.
\label{fig:hist}} 
\end{figure}

We also performed an exercise of applying the theory lying behind the
proper $\chi^2$ distribution. We used the best fit value of 
$\chi^2_{dist}(obs)$ from the Table~\ref{tab:psfit} and determined the error
contour by adding the standard value of 4.61 (Avni 1976) 
for 90\% error, two
parameters of interest (see Fig.~\ref{fig:chi2_46}). Such a contour is much
more compact than the contour determined directly from the simulations, 
as described above. This is surprising
since the histogram of $\chi^2_{dist}(i)$ approximates quite well the
theoretical $\chi^2$  histogram (see Fig.~\ref{fig:hist}) 
for the appropriate number of degrees of
freedom (22 in the case of X-ray data). Error contours determined
from simulations correspond to adding a value of order of  $\sim 8 - 28 $ 
to the best fit $\chi^2_{dist}(obs)$.  This shows that the Monte Carlo 
simulations are indeed extremely important, as argued by Uttley et al. (2002).

\section*{Appendix B: Structure function - error determination}

We again use two methods in order to determine the errors. The first method
is to assign directly the error of the structure function obtained
observationally. The second method includes the Monte Carlo simulations.  
Both methods are used in the present paper.

\subsection*{B.1 The direct method}

The structure function is computed in the timescale range determined by the
data. We use logarithmic time bins with a width of 0.16. We determine the
SF for bins with the number of pairs available greater than 7 pairs.  
The errors are assumed to come both from the flux measurement error and 
from the limited statistics of pairs. The flux measurement error is small, 
particularly for the photoelectric optical data (0.01, 0.015 and 
0.02 - 0.025 
mag for V, B and U band, correspondingly). The error on the data determined 
from the old photographic plates is below 20\% and the X-ray
data have errors of order of 10\%. Therefore, statistical errors dominate.

The statistical uncertainty in the SF$(\tau_k)$ is equal to the ratio of
the dispersion in the SF at a given bin to the number of pairs, i.e.

\begin{equation}
\sigma^{2 \quad (st)}_{SF} (\tau_k) = \sum_{ij} \frac{(x_i - x_j  -(\overline{x_i - x_j}))^2}{(n_k/2)^2} \ ,
\end{equation}
where $\overline{x_i - x_j}$ is the mean pair difference and $n_k$ is the
number of pairs in bin k. This error is plotted in Fig.~\ref{fig:SF}.
%~\ref{fig:sfa}.

In our error analysis we included two additional effects, which should 
render the error determination more robust. 

For completeness, we took into account the errors associated with 
the flux measurements.  This can be easy assessed through
Monte Carlo simulations.  Assuming that the errors in fluxes are normally
distributed, we can modify each flux by random Gaussian deviates
based on the quoted error for each data point. In a single Monte Carlo
realization each data point is modified, and the SF is computed.
Computing a large number of independent realizations (500 - 1000) we
obtain the mean SF and the rms deviation of the mean SF. We call this
rms deviation to result from flux randomization  
$\sigma^{2 \quad (fr)}_{SF} (\tau_k)$ (see Peterson, Wanders and Horne 1998).
The third source of errors is associated with the observational sampling
of the light curves. The specific times of observations are in a sense 
'randomly distributed' and the SF can be less sensitive to the choice of 
individual data points than the power spectrum. We can estimate these 
uncertainties through considering subsets of the original ``parent'' data set.
This method is known as the bootstrap method. It can be used to
evaluate the significance of the SF itself, its rms and other
parameters such as slope "{\it b}" of SF.  The method works as follows:  

We have a set of observations {(t1,x1), (t2,x2),..., (tn,xn)}.
We can take from this set n pairs of randomly selected points
without regard to whether or not they have been previously
selected.  Thus we select a subset of the original data points,
in which some data pairs repeat more than once and some of the original 
pairs are ignored. Next the multiple pairs are removed so
the size of the selected sample is reduced. For this data set we 
again calculate the SF and we repeat this process many times (500 - 1000).  
Multiple realizations lead to a mean and standard deviation for the
structure function based on a randomly chosen subset of the
original data points. We mark this rms value describing the effect of
random subset selection as  $\sigma^{2 \quad rss}_{SF} (\tau_k)$.

Our computations show that bootstrapping gives the most conservative
values of uncertainties in comparison with the other two described above. 
This is probably due to the fact that the considered subsets have 
smaller number of points than the real light curve.

The total uncertainty of the SF can be obtained if we assume that the 
errors add in quadrature.
\begin{equation}
\sigma^{2}_{SF}(\tau_k) = \sigma^{2 \quad (st)}_{SF}(\tau_k)+ \sigma^{2 \quad fr}_{SF}(\tau_k)+  \sigma^{2 \quad rss}_{SF}(\tau_k) \ .
\end{equation}

These error bars are used in Figs.~\ref{fig:sim7} and ~\ref{fig:sfa}.
% can see on the new Fig.2(c), Fig.5.  
This procedure allows us also to estimate the best-fit
value and uncertainties of such parameter as a slope $b$ of
linear part of the SF.

We note here that all those uncertainties are model independent.

\subsection*{B.2 Monte Carlo simulations of light curves}

In order to estimate the  errors of the SF without a significant loss
of information appearing in the bootstrap method we also perform the
Monte Carlo simulations of the light curve itself based on the specific
SF model.

The SF with the linear part having the slope $b$ can be obtained within 
the frame of the Poissonian model also called ``shot noise'' model.  
It means that in a physical system the variations are due to the 
stochastic superposition of independent flares of a given basic shape 
randomly distributed in time, but of various durations and
amplitudes.  Under certain conditions, imposed on the flare, the
amplitudes and durations of such process may result in the
first-order structure function $SF(\tau)$ being a power law of
$\tau$ with an index $b$ which takes on values from 0 to 2 (e.g. 
Terebizh et al. 1989; Sergeev 1999, Cid Fernandes, Sodr'e, \& Vieira
da Silva 2000).

Sergeev (1999) showed that if the number $n$ of flares of duration $T_f$ is
given by a power law distribution, i.e.
\begin{equation}
n(T_f) \sim T_f^\alpha
\label{eq:alpha}
\end{equation}
and if the flare amplitude $A(T_f)$ also depends on the flare duration
as a power law,
\begin{equation}
A(T_f) \sim T_f^\beta,
\label{eq:beta}
\end{equation}
then the dependence of the structure function on the interval $\tau$
asymptotically converges to a power law shape 
\begin{equation}
SF(\tau) \sim \tau^{b}, ~~~~~ b= \alpha+2\beta+2.
\end{equation}
In the considerations of Sergeev (1999), the extension of the power law
dependencies of the flare parameters given by Equations~\ref{eq:alpha} and 
\ref{eq:beta}
were limited by a natural physical 
constraints:  the flare duration was allowed to vary only between 
$T_f^{min}$ and $T_f^{max}$. The adopted shape of the flares was 
Gaussian, but it was shown that the profiles of flares are of no 
importance. Flares can have any profile:  square, triangle or 
Gaussian (e.g. Sergeev 1999, Cid Fernandes et al. 2000).

We performed Monte Carlo simulations using the software of Sergeev with
5 input free parameters: 

\begin{enumerate}

\item dT(flare)= the mean time interval between two flares
\item $\alpha$ = index in the dependence of flare numbers on duration,
\item $\beta$ = index in the dependence of the flare amplitude on duration,
\item $T_{min}$ = minimal duration of flares
\item $T_{max}$ = maximal duration of flares.

\end{enumerate}

Using Sergeev's program for simulation of light curves, 
we modeled such light curves using various values of $\alpha$ and $\beta$.
It is important to stress, however, that it is impossible to obtain
separately the values of the parameters $\alpha$ and $\beta$. Therefore,
we took into account the relation $ b = 2+\alpha + 2\beta$, where
$ b$ is a slope of the SF determined from observations.

Changing all 5 parameters we simulated a number of
light curves (N$\sim $500 - 1000) in each energy band and we
computed the SF for each simulated light curve in the same
manner as we did for the observed light curve.  
From each bin of all calculated SFs we subtracted 
$2D_{err}$ with $D_{err}$ being the mean variance of the measurement 
uncertainties.  The model SF (corresponding to the given model 
and sampling pattern of the observed light curve) is then given 
by the mean of the N simulated SFs, $\overline{SF(\tau)_{sim}}$.  
The model error is given by the rms spread of the simulated model 
population around the mean value at each time interval $\tau$, 
$\sigma(\tau)_{sim}$.

The quality of the model representation of the observational
light curves is estimated with help of statistics $\chi^2_{test}$,
which we calculate as follows:  
\begin{equation}
\chi^2_{test}=\sum_{T_f=T_{min}}^{T_f=T_{max}}
\frac{[\overline{SF(\tau)_{sim}}-SF(\tau)_{obs}]^2}
{\sigma ^2(\tau)_{sim}}
\end{equation}

We assumed that the best parameters of model are those yielding 
the minimum value of $\chi^2_{test}$ and the values of these 
parameters are listed in Table~\ref{tab:sffit}.

\ \\ 
This paper has been processed by the authors using the Blackwell 
Scientific Publications \LaTeX\  style file. 


\begin{thebibliography}{}

\bibitem []{} Abrassart A., Czerny B., 2000, A\& A, 356, 475 
%\bibitem []{} Aleksander T., Strurm E., Lutz D., Sterberg A., Netzer H., Genzel R., 1999, ApJ, 512, 204
\bibitem []{} Avni Y., 1976, ApJ, 210, 642
\bibitem []{} Belloni T., Mendez M., King A.R., van der Klis M., van Paradijs J., 1997, ApJ, 479, L145
\bibitem []{} Belokon E.T., Babadzanjantz M.K., Lyutyi V.M., 1978, A\&ASS, 31, 383 
\bibitem []{}Boller Th., Brandt W.N., Fabian A.C., Fink H.H., 1997, MNRAS, 289, 393 
\bibitem []{} Borczyk W., Schwarzenberg-Czerny A., Szkody, P., 2003, A\&A (submitted)
\bibitem []{} Brandt J.C., et al., 2001, AJ, 121, 2999
\bibitem []{} Brandt W.N.,  Boller Th., Fabian A.C., Ruszkowski M., 1999, MNRAS, 303, L53
\bibitem []{} Cannon R.D., Penston M.V., Brett R.A., 1971, MNRAS, 152, 79
%\bibitem []{} Chiang J., et al., 2000, ApJ, 528, 292
\bibitem []{} Cid Fernandes, R., Sodr'e Jr, L., Vieira da Silva, L., 2000,
ApJ, 544, 123
\bibitem []{} Collin-Souffrin S., Czerny B., Dumont A.-M., \. Zycki P.T., 1996, A\&A,  314, 393
\bibitem []{} Collier S., Peterson B.M., 2001, ApJ, 555, 775.
%\bibitem []{} Cui W.,  Zhang  S. N., Focke W., Swank J. H., 1997, ApJ, 484, 383 
%Transition state of Cyg X-1 
\bibitem []{} Czerny B., Lehto J.H., 1997, MNRAS, 285, 365
\bibitem []{} Czerny B., Niko\l ajuk M., Piasecki M., Kuraszkiewicz J., 2001, MNRAS, 325, 865
\bibitem []{} Czerny B., Schwarzenberg-Czerny A., Loska Z., 1999, MNRAS, 303, 148 
%\bibitem []{} De Vaucouleurs, G., De Vaucouleurs, A., 1968, AJ, 73, 638 
\bibitem []{} Doroshenko V.T., et al. 2001, AstL, 27, 65
%\bibitem []{} Doroshenko V.T., Lyutyi V.M., Shenavrin V.I., 1998, Astronomy Letters, 24, 160
%\bibitem []{} Done C., 2001, Adv. Space Res. (in press, astro-ph/0012380)  
\bibitem []{} Edelson R. et al. 1996, ApJ, 470, 364
%\bibitem []{} Edelson R., Krolik J.H., 1988, ApJ, 333, 646  
\bibitem []{} Evans I. N., Tsvetanov Z., Kriss G. A., Ford H. C., Caganoff S., Koratkar A. P., 1993, ApJ, 417, 82 
\bibitem []{} Fan J.-H., Su C.-Y.,  1999, ChA\&A, 23, 22
\bibitem []{} Fiore F., Massaro E., Perola G.C., Piro L., 1989, ApJ, 347, 171
\bibitem []{} Fitch W.S., Pacholczyk A.G., Weymann R.J., 1967, ApJ, 150, 67
\bibitem []{} Gierli\' nski M., Zdziarski A., Poutanen J., Coppi P.S., Ebisawa K., Johnson W.N., 1999, MNRAS, 309, 496
\bibitem []{} Gilfanov M., Churazov E., Revnivtsev M., 2000, MNRAS, 316, 923
\bibitem []{} Hayashida K., Miyamoto S.,  Kitamoto S.,  Negoro H.,  Inoue H., 1998, ApJ, 500, 642
%\bibitem []{} Hawley J.F., Krolik J.H., 2001, ApJ, 548, 348
\bibitem []{} Janiuk A., Czerny B., Siemiginowska A., 2000, ApJ, 542, L33
\bibitem []{} Janiuk A., Czerny B., Siemiginowska A. 2002, ApJ, 576, 908 
\bibitem []{} Johnson W.N., McNaron-Brown K., Kurfess J. D.,
Zdziarski A. A., Magdziarz P., Gehrels N., 1997, ApJ, 482, 173 
\bibitem []{} Kaspi S., et al. , 1996, ApJ, 470, 336 
\bibitem []{} Kataoka J., et al., 2001, ApJ, 560, 659
\bibitem []{} Kazanas D., Nayakshin S., 2001, ApJ, 550, 655
\bibitem []{} Kolmogorov A.N., 1941a, Dokl.Acad.Nauk.SSSR, v.30, p.229 
\bibitem []{} Kolmogorov A.N., 1941b, Dokl.Acad.Nauk.SSSR, v.32, p.19 % Reports of SU Academy of Science
\bibitem []{} Kriss G.A., Davidsen, A.F., Blair, W.P., Bowers, 
C.W. Dixon, W.V. et al. 1992, ApJ, 392, 485
\bibitem []{} Kuraszkiewicz J., Loska Z., Czerny B., 1997, Acta Astr., 47, 263  
\bibitem []{} Leighly K., 1999, ApJSS, 125, 297
\bibitem []{} Longo G., Vio R., Paura P., Provenzale A., Rifato A.,  1996, A\& A, 312, 424
\bibitem []{} Lyuty V.M., Doroshenko V.T., 1999,  Astron. Lett., 25, 341 
\bibitem []{} Lyutyj V.M., Oknyanskij V.L., Chuvaev K.K. 1984, 
Sov. Astron. Let., 10, 335 
\bibitem []{} Lyutyj V.M., Oknyanskij, V.L. 1987 Astron.Zhurnal, 64, 
465 
\bibitem []{} Lyutyi V.M.,Taranova, O.G., Shenavrin, V.I. 1998 
Astron. Lett., 24, 199
\bibitem []{} Manners J., Almaini O., Lawrence A., 2001, MNRAS, 330, 390 
\bibitem []{} Markowitz, A., Edelson, R., 2001, ApJ, 547, 684
\bibitem []{} McHardy I.M., Czerny B., 1987, Nat., 325, 696
\bibitem []{} Merkulova N.I., Metik L.P., Pronik I.I., 2001, A\&A, 374, 770
\bibitem []{} Mushotzky R.F., Done C., Pounds K.A., 1993, ARA\&A, 31, 717
\bibitem []{} Ogle P.M., Marshall H.L., Lee J.C.,
 Canizares C. R., 2000, ApJ, 545, L81 
\bibitem []{} Oknyanskiy V.L., 1978, Perem. Zvezdy, 21, 71 
\bibitem []{} Oknyanskiy V.L., 1983, Astron.Circuliar, 1300, 1 
\bibitem []{}  Oknyanskij V.L., 1994,  Astrophysics and Space Science, 222, 157
\bibitem []{} Oknyanskij V.L., Lyuty V.M., Taranova O.G., Shenavrin V.I., 1999, Astronomy Letters, 25, 483
\bibitem []{} Pacholczyk A.G., 1971, ApJ, 163, 449 
\bibitem []{} Pacholczyk A.G., et al., 1983, Astroph. Letters, 1983, 23, 225
\bibitem []{}  Paltani S., Courvoisier T.J.-P.,  Walter R., 1998, A\&A, 340, 47
\bibitem []{} Papadakis I., McHardy I.M., 1995, MNRAS, 273, 923
\bibitem []{} Papadakis I., Lawrence A., 1993, MNRAS, 261, 612
\bibitem []{} Penston M.V., P\'erez E. 1984, MNRAS, 211, 84 
\bibitem []{} Penston M.V., Penston, M. J., Neugebauer G., Tritton K. P.,
 Becklin, E. E., Visvanathan, N., 1971, MNRAS, 153, 29
\bibitem []{} Perez G. et al. 1998, ApJ, 500, 685 
\bibitem []{} Perotti F. et al., 1990, ApJ, 356, 467
\bibitem []{} Peterson B.M., Wandel A., 1999, ApJ, 521, L95
\bibitem []{} Peterson B.M., Wanders I., Horne K., 1998, PASP, 110, 660
\bibitem []{} Pottschmidt K., et al., 2002, A\& A (astro-ph/0202258)
\bibitem []{} Press W.H., Rybicki G.B. 1997, in Astronomical Time 
       Series, ed. D. Maoz, A. Sternberg, \& E. M. Leibowitz
       Dordrecht, Kluwer, p. 61 
\bibitem []{} Pronik S. G., Sergeev V.I., Sergeeva E. A., 2001, ApJ, 554, 245
\bibitem []{} Reina C., Tarenghi M., 1973, A\&A, 26, 257
\bibitem []{} Risaliti G., Elvis M., Nicastro F., 2002, ApJ, 571, 234
%\bibitem []{} Roberts, D.H., Leh\' ar J., Drehner J.W., 1987, AJ, 93, 968
\bibitem []{} Rokaki E., Collin-Souffrin S., Magnan C., 1993, A\&A, 272, 8
\bibitem []{} R\' o\. za\' nska A., Czerny B., \. Zycki P.T., Pojma\' nski G., 1999,
    MNRAS, 305, 481
\bibitem []{} Sergeev S.G., 1999, Candidate's dissertation,
   Krymsk. Astrofiz. Obs., Crimea
\bibitem []{} Sergeev S.G., Pronik V.I., Sergeeva E.A., Malkov Yu.F., 1999,ApJSS,
  121, 159
\bibitem []{} Shakura N.I., Sunyaev R.A., 1973, A\&A, 24, 337 
\bibitem []{} Siemiginowska A., Czerny B., 1989, MNRAS, 239, 289
\bibitem []{} Simonetti J.H., Cordes J.M., Heeschen D.S., 1985, ApJ 1985, 296, 46 
\bibitem []{} Szuszkiewicz, E., 1999, Mem.S.A.It, 70, 95 
\bibitem []{} Terebizh V.Yu.,  Terebizh A.V. \& Biryukov V.V., 1989, Astrofizika, 31, 75
\bibitem []{} Timmer J., K\" onig M., 1995, A\&A, 300, 707
\bibitem []{} Ulrich M.-H., 2000, A\&ARv, 10, 135
\bibitem []{} Ulrich M.-H., Maraschi L., Urry M.C., 1997, ARA\&A, 35, 445 
\bibitem []{} Uttley P., McHardy I.M., 2001, MNRAS, 323, L26
\bibitem []{} Uttley P., McHardy I.M., Papadakis I.E., 2002, MNRAS, 332,
231 
\bibitem []{} Wandel A., Peterson B.M.,  Malkan M.A., 1999, ApJ, 526, 579
\bibitem []{} Wang J.-X., Wang T.-G., Zhou Y.-Y., 2001, ApJ, 549, 891 
\bibitem []{} Ward M.J., Geballe T., Smith M., Wade R., 
Williams P., 1987, ApJ, 316, 138 
\bibitem []{} Weaver K.A., Yaqoob T., Holt S.S.,
 Mushotzky R.F., Matsuoka M.,
 Yamauchi M., 1994, ApJ, 436, L27 
\bibitem []{} Wu Ch.-Ch., Weedman D.W. 1978, ApJ, 223, 798 
\bibitem []{} Yaqoob T., et al, 1993, MNRAS, 262, 435 
\bibitem []{} Yaqoob T., Warwick R.S., 1991, MNRAS, 248, 773 
\bibitem []{} Zdziarski A.A., Johnson W.N., Magdziarz P., 1996, MNRAS, 
    283, 193
\bibitem []{} \. Zycki P.T., R\' o\. za\' nska A., 2001, MNRAS, 325, 197
 
\end{thebibliography}
\end{document}